\pdfoutput=1
\documentclass[numbers]{article}
\usepackage{arxiv}

\usepackage[utf8]{inputenc} %
\usepackage[T1]{fontenc}    %
\usepackage{hyperref}       %
\usepackage{url}            %
\usepackage{booktabs}       %
\usepackage{amsfonts}       %
\usepackage{xspace}
\usepackage{dsfont}         %
\usepackage{nicefrac}       %
\usepackage{microtype}      %
\usepackage{amsmath}         %
\usepackage{cleveref}       %
\usepackage{graphicx}
\usepackage[font=scriptsize, labelfont=bf]{caption}
\usepackage{lineno}
\usepackage[numbers,sort&compress]{natbib}
\usepackage{algorithm}
\usepackage{algpseudocode}
\usepackage{multirow}

\usepackage{doi}
\usepackage[separate-uncertainty=true]{siunitx}    %

\title{Data-driven simulation of the brain's response to stimulation: a system identification approach for closed-loop neuromodulation, piloted in adaptive deep brain stimulation}

\usepackage{authblk}

\newbox{\orcid}\sbox{\orcid}{\includegraphics[scale=0.06]{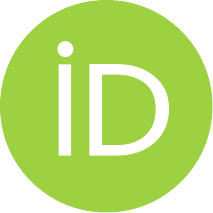}}
\author[1,2,3]{%
	\href{https://orcid.org/0009-0003-1477-4912}{\usebox{\orcid}\hspace{1mm}Matthias Dold\thanks{\texttt{matthias.dold@donders.ru.nl}}}%
}
\author[3, 4, 1]{%
	\href{https://orcid.org/0000-0002-2032-8981}{\usebox{\orcid}\hspace{1mm}Joana Pereira\setcounter{footnote}{0}}%
}
\author[2]{%
	\href{https://orcid.org/0000-0003-0030-7224}{\usebox{\orcid}\hspace{1mm}Marcus L.F. Janssen}%
}
\author[1]{%
	\href{https://orcid.org/0000-0001-6729-0290}{\usebox{\orcid}\hspace{1mm}Michael Tangermann\setcounter{footnote}{0}\thanks{\texttt{michael.tangermann@donders.ru.nl}}}%
}

\affil[1]{Data-Driven Neurotechnology Lab, Donders Institute for Brain, Cognition and Behaviour, Radboud University, Nijmegen, The Netherlands}
\affil[2]{Department of Clinical Neurophysiology, Maastricht University Medical Center, The Netherlands;  Mental Health and Neuroscience Research Institute, Maastricht University, The Netherlands}
\affil[3]{Department of Stereotactic and Functional Neurosurgery, University of Freiburg - Medical Center, Breisacher Straße 64, 79106, Freiburg, Germany}
\affil[4]{Brain-Links Brain-Tools, Intelligent Machine-Brain Interfacing Technology (IMBIT), University of Freiburg, Georges-Köhler-Allee 201, 79110, Freiburg, Germany
}

\renewcommand{\shorttitle}{Data-driven simulation of signal response to DBS}

\hypersetup{
pdftitle={E-SINDy for aDBS},
pdfsubject={q-bio.NC, q-bio.QM},
pdfauthor={Matthias Dold, Joana Pereira, Marcus L.F. Janssen, Michael Tangermann},
pdfkeywords={DBS, EEG, system identification, brain modeling, neural decoding, Parkinson's Disease},
}

\newcommand{\Castano}{Casta{\~n}o-Candamil }
\newcommand{\handmotor}{hand-motor }

\newcommand{\indicator}{\mathds{1}}

\newcommand{\ON}{\textsc{ON}\xspace}
\newcommand{\OFF}{\textsc{OFF}\xspace}

\begin{document}\maketitle

\begin{abstract}

\textbf{Introduction}--Closed-loop neuromodulation uses a physiomarker to control stimulation. Developing controllers based on physiomarker requires exploring a large space of strategies and parameters, which is often not feasible with the human-in-the-loop. Simulation environments are a possible alternative to close this gap. In Parkinson's disease, adaptive deep brain stimulation (aDBS) is a stereotypical application facing this constraint: while physiomarker identification is progressing rapidly, clinically available control is limited to threshold and proportional strategies.

\textbf{Methods}--We propose to use system identification within the sparse identification of nonlinear dynamics (SINDy) framework to create dynamical models of the brain with and without DBS.
Identified ex post from existing recordings, such models are a middle ground between abstract bottom-up models, e.g., spiking neuron networks or neural mass models, and the realistic but difficult-to-access human-in-the-loop.
We provide a proof-of-concept with EEG data from a single subject using SINDy models to forecast neural activity from real data initial conditions, thereby creating a simulation environment, which enables to answer the question \textit{what if} deep brain stimulation (DBS) is changed? Single threshold, proportional, proportional-integral-derivative (PID), and proximal policy optimization (PPO) control are trained and evaluated within this simulation environment. The objective is to have the simulated data classified as DBS \ON based on its signal characteristics, while minimizing the energy consumption to achieve this.

\textbf{Results}--It is possible to fit SINDy models on spatially filtered EEG data while maintaining DBS condition-specific characteristics. All aDBS controllers trained in the simulation environment provide DBS more efficiently than continuous DBS (cDBS) within the same 2\,s simulation window, based on our reward calibration. Efficiency gains in terms of reduced total energy achieve up to $20.2\,\%$ reduction compared to cDBS. Proportional control yields a small but significant gain over threshold control. PID and PPO achieve a comparable reward advantage over cDBS, with PPO stimulating more conservatively and PID saving the most energy.

\textbf{Discussion}--The proposed approach of using system identification to create a simulation environment has proven feasible and allowed comparison of different optimized controllers. Within this data-driven simulation environment, PID control is a strong a-priori choice, matching the reward of deep reinforcement learning at a lower energy cost, while the benefit of more complex reinforcement learning control may only emerge with richer physiomarker signals.

\end{abstract}

\keywords{DBS \and aDBS \and EEG \and neural markers \and system identification \and brain modeling \and closed-loop neuromodulation}

\section{Introduction}

Closed-loop neuromodulation adapts stimulation based on a physiomarker which is informative about the need for stimulation. Examples range from adaptive deep brain stimulation (aDBS)~\cite{Little2013, Neumann2023, DeNeeling2026} over responsive neurostimulation in epilepsy~\cite{Morrell2011, Nair2020} to closed-loop spinal cord stimulation~\cite{Mekhail2019, Milekovic2023, Lorach2023}. Common to all of these is the need for a control strategy (\textit{controller}), which decides how to stimulate given the physiomarker. Designing and tuning such controllers requires exploring a large space of strategies and parameters, which is often not feasible with a human-in-the-loop setup, as experimental time with patients is limited and each evaluation is costly.

ADBS for Parkinso's disease (PD) is a stereotypical neuromodulation application facing the same constraint, and is the domain the data for this contribution originates from.

On a high level, an aDBS system comprises three components: 1) a sensor with (near) real-time recording and streaming capabilities, 2) a decoding algorithm which extracts a physiomarker from sensor data, 3) a control algorithm which decides how to stimulate based on the physiomarker's current state, and a neuro-stimulator which allows (near) real-time adjustment of the stimulation parameters. Steps 1) and the neuro-stimulator of step 3) are mainly driven by hardware advances, the others are predominantly algorithmic/software bound. The algorithmic development in identifying physiomarkers is progressing quickly and producing various candidates~\cite{Little2013, Tinkhauser2017, Olaru2025, Castano2020, Stam2025}, converging to the need for patient specific evaluation~\cite{Gerster2025, deNeeling2026b}. The development of control strategies which make it to the final evaluation with a human-in-the-loop is progressing more slowly. Central challenges are the limited interaction time when applying DBS on externalized leads, or the limited hardware and streaming capabilities when working on implanted systems.
From a clinical perspective~\cite{DeNeeling2026}, aDBS control is mainly associated with threshold~\cite{brontestewart2025, Little2013} and proportional control~\cite{Rosa2015, Arlotti2021}, as these are available on certified aDBS-ready implantable pulse generator (IPG) from Medtronic (Minneapolis, MN, USA) and Newronika (Milan, Italy).

Computational models on the other hand provide a more accessible but less realistic environment for investigating and designing aDBS controllers~\cite{Kuzmina2025}. These range from threshold based~\cite{Fleming2023}, over proportional-integral-differential (PID)~\cite{Gorzelic2013, Su2019, Fleming2023} and analogous control approaches~\cite{Zhu2021}, to reinforcement learning (RL) controllers such as proximal policy optimization (PPO)~\cite{Gao2026synapse, Krylov2020}, or general actor-critic based approaches~\cite{Gao2023}. The underlying simulation environments are often based on a basal-ganglia-thalamo-cortical model first introduced by Rubin and Terman~\cite{Rubin2004} (extension of the Hodgkin-Huxley model for DBS), or firing-rate based~\cite{Polli2025} and neural mass models~\cite{West2022}.
Such bottom up models of Parkinsonian dynamics create a rich evaluation environment for training these partially more complex control strategies. They are however bound to inherent design choices capturing various stages of realism~\cite{Kuzmina2025}.

A middle ground between the bottom up and human-in-the-loop models are data-driven approaches, which identify a model of the response to stimulation from recorded data and subsequently design or tune a controller~\cite{Acharya2022}. Most of these models are linear and treat stimulation as an exogenous input directly in the model. Haddock et al.~\cite{Haddock2017} fitted autoregressive models with exogenous input (ARX models) of the tremor band power measured with an inertial measurement unit under DBS \ON and \OFF conditions. The ARX models were then used to optimize a model predictive controller. Yang et al.~\cite{Yang2018} proposed a system identification framework based on linear state-space models (LSSM), and identified a stochastic stimulation pattern for an optimal identification of the input-output system. The framework was later applied to local field potentials (LFP) of non-human primates, where model-based closed-loop control~\cite{Yang2021} was then simulated. Wang et al.~\cite{Wang2017} used existing LFP data from rodents to fit an ARX model with DBS \ON/\OFF as input on a theta-band marker, and evaluated a closed-loop controller. These approaches either require dedicated experiments (specific stimulation sequences), are based on animal data, or model behavioral instead of neural signals. In all cases, the models enabled a simulation environment for evaluating the controllers.

With this work, we propose to derive such a simulation environment ex post from human electroencephalogram (EEG) data, which were recorded under different DBS conditions. Instead of treating DBS as an exogenous input, we use system identification~\cite{Brunton2019Book} to model the neural signal dynamics within each DBS condition. Like the approaches above, it is a middle ground between using the actual human-in-the-loop system and the abstract computational bottom-up models. Specifically the latter are often tuned to produce group level phenomena, such as beta synchrony~\cite{Kuehn2008}, that might not be present in the individual~\cite{Gerster2025} or might not be susceptible to DBS in individual cases~\cite{Dold2026}. Using a data-driven system identification approach further removes the need to hand-select phenomena that should be integrated into the simulation model~\cite{Kuzmina2025}.

We apply this approach on a proof-of-concept, using a regularized form of Sparse Identification of Nonlinear Dynamics (SINDy)~\cite{Brunton2016, Fasel2022} on data of a single patient from a previously published EEG dataset ~\cite{Castano2020, Dold2026, dold2026dataset}. Combined with preprocessing for an optimized DBS contrast, we create models describing the inherent dynamics of the DBS \ON and \OFF regimes. The models are used with numeric integration to build a novel simulation environment in which we optimize and evaluate \textit{threshold}, \textit{proportional}, proportional-integral-derivative (PID) and proximal policy optimization (PPO) based aDBS control. While this proof-of-concept is specifically done in the domain of aDBS, the approach in and of itself is more generally applicable for any offline data set with recordings of at least two different conditions.

\section{Methods}

\begin{figure}
    \centering
    \includegraphics[width=\linewidth]{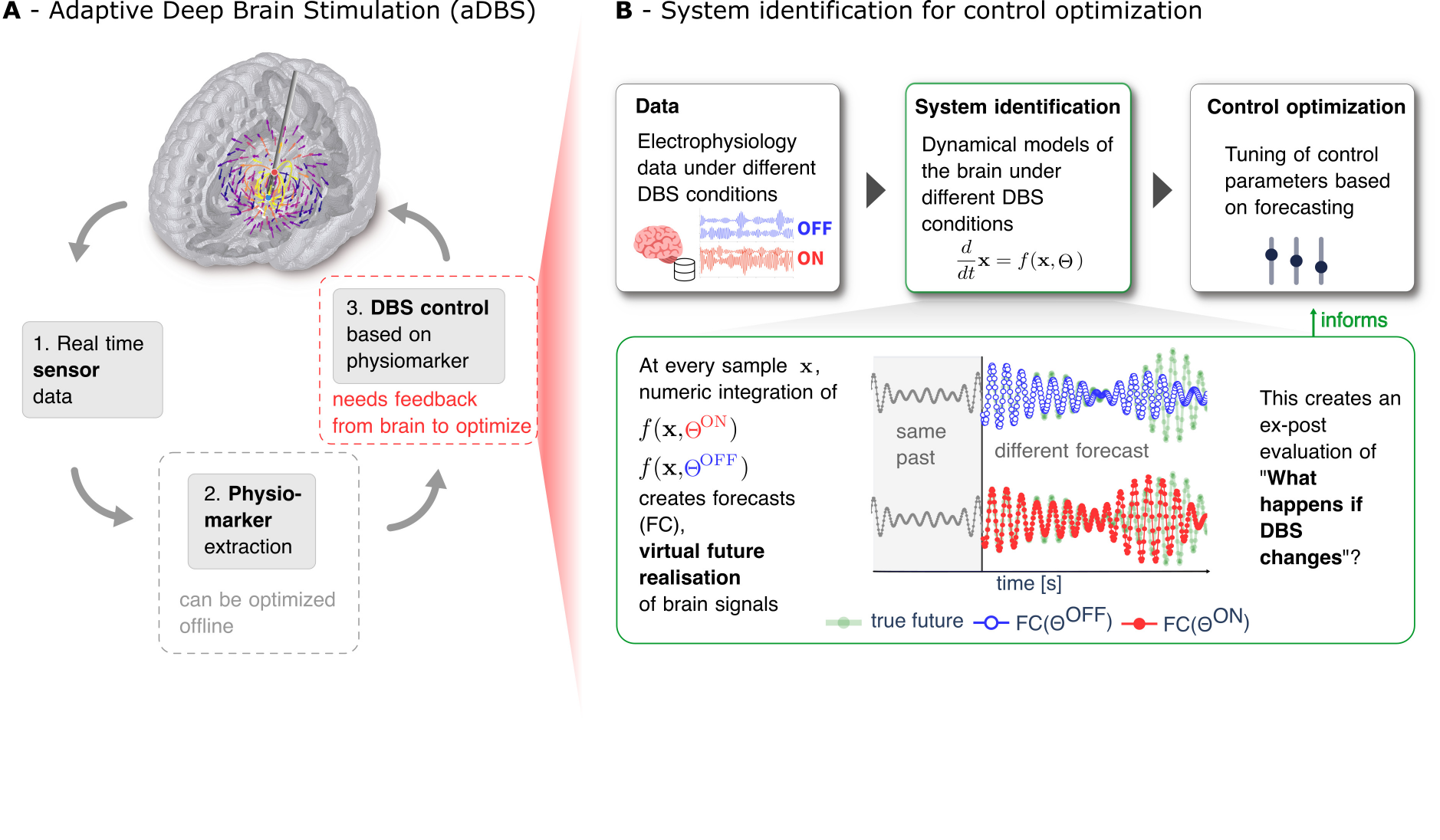}
    \caption{System identification for brain stimulation control optimization. Panel \textbf{A} illustrates the basic components of an aDBS system, containing streaming sensor data (electrophysiology or other), a physiomarker extraction or decoding algorithm, and a DBS control strategy based on the physiomarker. While the physiomarker can be identified from offline data, optimizing the control strategy requires feedback from the system it tries to control, i.e., the brain. Panel \textbf{B} is a schematic of the proposed approach to use preexisting DBS labeled data to derive dynamical models of the brains under various DBS conditions. These models can then be used to forecast brain activity, under different DBS conditions, creating a hypothetical future or simulation. The simulation can be used to evaluate control strategies which change the stimulation during the forecasting.}
    \label{fig:approach_schema}
\end{figure}

\begin{figure}
    \centering
    \includegraphics[width=\linewidth]{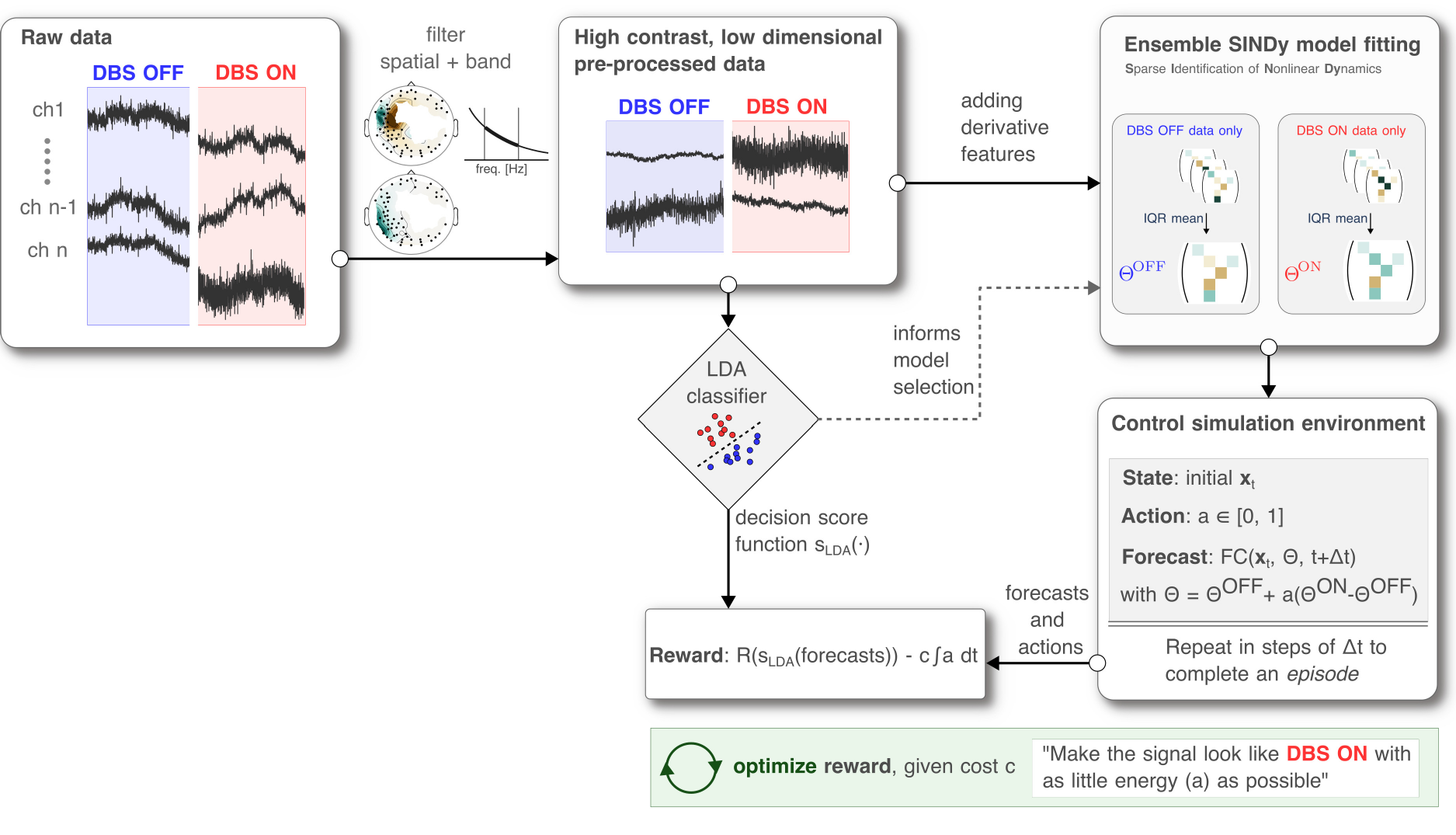}
    \caption{Full schematic of the data-driven sparse identification of nonlinear dynamics (SINDy) for optimization of adaptive deep brain stimulation (aDBS) control. The flow starts from any raw data, recorded from two different DBS conditions. In our proof-of-concept, EEG data were recorded while DBS was ON or OFF, in alternating blocks. Preprocessing and dimensionality reduction consisted in common spatial patterns and frequency-band filtering, resulting in two channels with optimized contrast for the DBS conditions. Derivative features are added by numeric differentiation, and ensemble SINDy models are trained on epoched data individually for each DBS condition. Additionally, a linear discriminant analysis (LDA) classifier is trained to identify the DBS condition of individual data segments of the pre-processed two dimensional time series. The SINDy models are used within a simulation environment where an affine combination based on the action or energy parameter, a, is used to create a model of mixed dynamics. A control strategy can change this parameter every $\Delta t=250\,\text{ms}$ based on input from the past observed data. An episode of simulation (2\,s) is then evaluated by the LDA and a reward is calculated by subtracting the total energy times a cost factor, c. This simulation of brain activity can be used to optimize a controller, based on the following objective: make the episode's (here, a 2\,s window) be classified as DBS ON---"look like DBS ON"---with as little energy expenditure as possible, under the assumption that the DBS ON state is the better state for the patient.}
    \label{fig:approach_detail_flowchart}
\end{figure}

The core idea of our approach lends itself from the good regulator theorem~\cite{Conant1970}. It states that each good regulator (or control algorithm), is also a good model of the system (the brain). A model of the system and a good controller are hence inherently linked and we therefore propose to first create a model of the system, which we then use to build a good controller. In our case, the model is a (reduced) abstraction of the brain processes, derived from a patients brain recordings. In the following, we describe how we use existing EEG data, derive models describing the dynamics of the observed data under DBS ON and OFF, and finally use these models in a simulation environment to train various control strategies. Figure~\ref{fig:approach_schema} illustrates the idea of the approach on a high-level, while Figure~\ref{fig:approach_detail_flowchart} provides the experiment flow for the provided proof-of-concept.

\subsection{Data}
For the proof-of-concept, EEG signals are used from a session of a single PD patient receiving DBS on externalized leads in a post-surgical sub-acute phase. We chose data of the session that was found to have the strongest behavioral DBS effect observed in the cohort (S2, session 2)~\cite{Dold2026}. EEG signals from 64 channels were recorded at 1~kHz sampling rate using BrainAmp DC amplifiers (Brain Products GmbH, Gilching, Germany), while the participant performed the CopyDraw~\cite{Castano2019} task under alternating DBS conditions. Details about the CopyDraw task are not relevant for this work but can be found in \Castano et al.~\cite{Castano2019}. In short, the CopyDraw task engages the participant in a \handmotor copy-drawing of a target trace using a digital stylus. One repetition of the task is referred to as a \textit{trial}. Twelve such trials are grouped within a \textit{block}. DBS was constant within each block (either \ON or \OFF) and switched in an alternating pattern, resulting in an equal number of DBS \ON and \OFF blocks. The total data set for the sample patient contains twelve blocks (six DBS \ON and six DBS \OFF) with twelve trials each. A single trial lasted six seconds. Preprocessing contained visual inspection and exclusion of bad channels, ICA artifact removal as outlined in \Castano et al.~\cite{Castano2019}, and resampling to 300~Hz.

\subsection{Dimensionality reduction and signal-to-noise optimization}
\label{sec:dim_red}
\begin{figure}
    \centering
    \includegraphics[width=\linewidth]{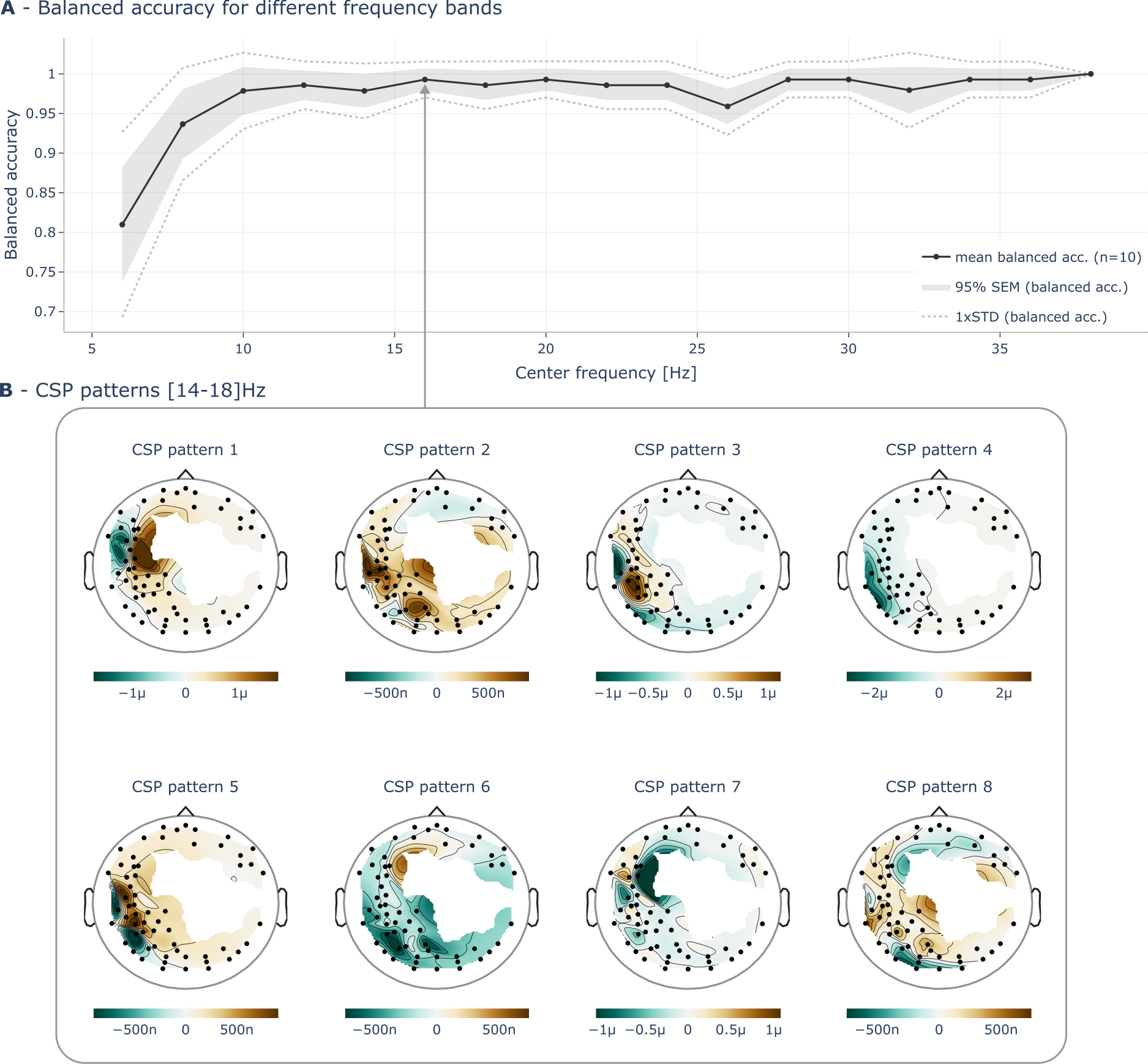}
    \caption{Grid search and common spatial patterns (CSP). \textbf{A} - shows a grid search over multiple frequency bands of 4~Hz width with band centers on the x-axis. The y-axis shows the balanced accuracy of a classifier pipeline using common spatial patterns (CSP) log-bandpower and a shrinkage linear discriminant analysis to predict the DBS condition from band pass filtered data. The dark trace with markers shows the mean balanced accuracy across a ten fold shuffled cross-validation. Gray shaded area and dashed lines represent the standard error of the mean (SEM) and $\pm1$ standard deviation (STD) bands. \textbf{B} - shows the CSP spatial patterns for the selected frequency band of 14--18~Hz.}
    \label{xfig:grid_search_and_csp_patterns}
\end{figure}

Motivated by the machine learning decoding approach used on the same data~\cite{Castano2020, Dold2026}, we use regularized common spatial patterns (CSP)~\cite{Lotte2010} in combination with shrinkage linear discriminant analysis (LDA, Ledoit-Wolf shrinkage)~\cite{Ledoit2004} to identify CSP patterns in a narrow frequency band, which are maximally informative about the DBS condition. For this purpose, a grid search with frequency bands of 4~Hz width is applied over a range of 4--40~Hz with 2~Hz overlap. For each narrow band, CSP are fitted and eight components selected, four from each end of the eigenvalue spectrum. The log-bandpower of each spatial filter was then used to classify the DBS condition with LDA. Looking at the mean balanced accuracy per frequency band center (appendix Figure~\ref{xfig:grid_search_and_csp_patterns} A), we chose the 14--18~Hz band as a lower beta frequency band with almost perfect balanced accuracy ($99.3\,\%$)--CSP spatial patterns of the selected band are shown in the appendix Figure~\ref{xfig:grid_search_and_csp_patterns} B. Targeting beta band activity is furthermore the default choice in aDBS~\cite{Gerster2025, he:2023, brontestewart2025}.
If $\mathbf{x}_t \in \mathbb{R}^{C_{EEG}}$ is the narrow-band time series of the $C_{EEG}$ EEG channels with time index $t$, we can use the CSP filters $\textbf{w}_{\text{CSP}_i}\in \mathbb{R}^{C_{EEG}}$ to get the CSP source time series of the $i$th CSP source by linear projection $\text{CSP}_{i, t}:=\textbf{w}_{\text{CSP}_i}^\top\cdot\mathbf{x}_t$. The time index is dropped to simplify the notation whenever the context makes clear that a time series is considered, and we refer to these time series as CSP channels. By plotting the bivariate trajectory of each pair of spatial filters (see appendix Figure~\ref{xfig:csp_phase_space}, one trajectory per trial), we selected two channels ($\text{CSP}_1$, $\text{CSP}_4$) which visually showed the clearest and most stable separation in the trajectories under DBS \ON vs.~\OFF. This reduced data to two channels with an optimized variance difference. The LDA was retrained on just these two channels and is used throughout the process whenever a classification is required.

\subsection{Modelling electrophysiology of the brain}
\subsubsection{Robust sparse identification of nonlinear systems for electrophysiology data}

While there are various approaches for modeling brain dynamics in the DBS context~\cite{Rubin2004, Fleming2020, Polli2025, West2022, Kuzmina2025}, we propose using the data-driven sparse identification of nonlinear systems (SINDy)~\cite{Brunton2016}. Using a data-driven system identification, is a middle ground between the true biological model, i.e., a human-in-the-loop, and a bottom-up model, e.g., spiking neural models, motivated by biological observations~\cite{Kuzmina2025}. Compared to bottom-up computations, the SINDy approach faces subject-specificity and noisy data modalities, which are challenging but also reflect real-world conditions.

So far, system identification has been used in the DBS context mostly for linearizing brain dynamics~\cite{Acharya2022}, deriving linear time-invariant systems with stimulation as exogenous input, which are then used to find control strategies, e.g., model predictive control (MPC), linear quadratic regulation (LQR), or fuzzy control~\cite{Yang2018, Yang2021, Wang2017, Ahmadipour2021, Liang2023, Kia2025}. Our approach is different in that it identifies two potentially nonlinear models of brain dynamics under DBS \ON and \OFF.
This is similar to Haddock et al.~\cite{Haddock2017}, who fitted two linear ARX models of the tremor band power of inertial sensor data instead of neural signals.
Given the two models, we then use the vector fields of both to model simulate brain dynamics as a continuous mixture of both. This has the advantage that we do not require the DBS action to be additive and state-independent on an otherwise fixed dynamical system. We are not confining the dynamics to a linearized system around an operating point. Furthermore, our approach can be realized with any neural signal recorded under two labelled conditions. Specifically, it does not require observations of the brain on specific predefined stimulation sequences~\cite{Yang2018, Yang2021, Haddock2017} or other specific experimental setups. SINDy has previously been applied in the context of DBS for depression to characterize stimulation-evoked oscillations~\cite{Tiruvadi2022}, but, to our knowledge, not to build a simulation environment for control.

The SINDy~\cite{Brunton2016} algorithm frames model identification as a sparse regression problem. To capture dynamics of ordinary differential equations (ODEs) of order $k$, we represent the system in state-space form as a system of first-order ODEs.

Given a recorded multi-channel signal $\mathbf{S}=[\mathbf{s}_1, \mathbf{s}_2, \ldots, \mathbf{s}_C]\in\mathbb{R}^{T\times C}$, where $\mathbf{s}_j$ is a time series of $T$ samples for channels $j\in \{1,2,\ldots, C\}$, we define the extended data matrix $\mathbf{X}$ by stacking the signals and their time derivatives up to order $k-1$:

\begin{equation}
    \mathbf{X} = \left[ \mathbf{S},\, \mathbf{S}^{(1)},\, \mathbf{S}^{(2)},\, \dots,\, \mathbf{S}^{(k-1)} \right] \in \mathbb{R}^{T \times (C \cdot k)}\,,
\end{equation}
where $\mathbf{S}^{(i)}$ denotes the $i$-th derivative with respect to time. Column $j$ of $\mathbf{X}$ is referred to by $\mathbf{x}_j$. The $k^{th}$ order system of linear ODEs is then given by

\begin{equation}
\label{eq:linearODE}
\dot{\mathbf{X}} =\mathbf{X}\cdot\mathbf{\Theta}_{\text{lin}}\,,
\end{equation}

with the real coefficient matrix $\mathbf{\Theta}_{\text{lin}}\in\mathbb{R}^{(C\cdot k)\times (C\cdot k)}$.

Extending the right hand side of equation~\ref{eq:linearODE} to consider nonlinear terms---often polynomials or trigonometric functions in the SINDy context---is achieved by using a library $\Lambda(\mathbf{X})$ of $N$ candidate terms from the columns of $\mathbf{X}$, e.g.,

\begin{equation}
\Lambda(\mathbf{X})= \left[1,\,\mathbf{X},\,\mathbf{X}^{\text{P2}},\,\mathbf{X}^{\text{P3}},\,\ldots,\,\sin(\mathbf{X}),\,\cos(\mathbf{X}),\,\ldots\right]\in\mathbb{R}^{T\times N}.
\end{equation}

Here, $\mathbf{X}^{\text{P2}}$ refers to all quadratic terms row-wise of $\mathbf{X}$, i.e., all unique monomials of order two

\begin{equation}
\mathbf{X}^{\text{P2}} = \left[\mathbf{x}_1^2,\,\mathbf{x}_1\mathbf{x}_2,\,\ldots,\,\mathbf{x}_{C\cdot k}^2\right].
\end{equation}

Similarly, $\mathbf{X}^{\text{P3}}$ would refer to all third order monomials, $\sin(\mathbf{X})$ to row-wise sine transforms ($\left[\sin(\mathbf{x}_1),\,\sin(\mathbf{x}_2),\,\ldots\right]$), and so forth. Given a suitable library, a system of nonlinear ODEs can then be formulated as

\begin{equation}
\label{eq:SINDyODENoNoise}
\dot{\mathbf{X}} = (\Lambda(\mathbf{X})\cdot\mathbf{\Theta}) \in \mathbb{R}^{T\times (C\cdot k)}\,,
\end{equation}

with coefficients $\mathbf{\Theta}\in\mathbb{R}^{N\times (C\cdot k)}$.

Based on the observation that many physical systems rely on a limited set of dominant mechanisms, $\mathbf{\Theta}$ is assumed to be sparsely populated in the SINDy context~\cite{Brunton2016} and the system is identified by solving Eq.\ref{eq:SINDyODENoNoise} in a sparsity promoting regularized linear least squares optimization

\begin{equation}
    \mathbf{\Theta}^{*} = \arg\min_{\mathbf{\Theta}} \|\dot{\mathbf{X}} - \Lambda(\mathbf{X})\mathbf{\Theta}\|_2^2 + \mathbf{R}(\mathbf{\Theta})\,,
\end{equation}

where $\mathbf{R}(\mathbf{\Theta})$ is a general regularization function. While multiple other forms of regularization~\cite{Zheng2019, Dong2023, Basiri2025} have been used successfully with the SINDy algorithm, we stick to the initially proposed sequential thresholded least squares (STLSQ)~\cite{Brunton2016} for efficiency reasons. Furthermore, convergence is ensured under relatively mild conditions~\cite{Zhang2019} for STLSQ, which requires only a single threshold hyperparameter to define the desired sparsity of the coefficient matrix $\mathbf{\Theta}$, keeping the hyperparameter space lean.

To address the often noisy and non-stationary signal encountered in electrophysiology, we make use of the robust Ensemble-SINDy (E-SINDy) framework~\cite{Fasel2022}. E-SINDy builds an ensemble of $M$ candidate coefficient matrices, $\{\mathbf{\Theta}^{m}\}_{m=1}^M$, commonly by bagging~\cite{Breiman1996} or library subsampling. To aggregate these into a final robust matrix $\mathbf{\Theta}$, an empirical inclusion probability is calculated for each entry $\theta_{ij}$ as the fraction of ensemble members with $\theta_{ij}^{m} \neq 0$; entries below a predefined threshold are set to zero, which preserves sparsity.
Instead of resampling, we use the natural segmentation of our data in trials and fit one SINDy model per trial, i.e., $M=72$ ensemble members per DBS condition (0--4\,s of each trial).
We then used an inclusion probability threshold of 0.5 (a coefficient needs to be present in 50\,\% of the individual models), and aggregated the coefficients using the inter-quartile mean (IQR mean), i.e., the mean over the ensemble members whose value lies between the first and third quartile, to further stabilize the ensemble models.

Despite these steps for robust models, stability for numeric integration is not ensured. A stable linear model is characterized by all eigenvalues $\lambda_j\in \mathbb{C}$ of $\mathbf{\Theta}_{\text{lin}}$ having $\text{Re}(\lambda_j)<0$ (real part less than zero). Nonlinear terms can stabilize an otherwise unstable linear core, but adding the sparsity constraints and the regularization steps can lead to eradicating damping provided by such terms. To ensure a certain level of stability, we propose to clip $\text{Re}(\lambda_j)\leq-\epsilon$, where $\epsilon$ is another hyperparameter. To modify only eigenvalues violating this bound and only their real parts, we apply clipping in the Schur decomposition $\mathbf{\Theta}_{lin} = \mathbf{Q} \mathbf{U} \mathbf{Q}^\top$,
where $\mathbf{Q}$ is orthogonal and $\mathbf{U}$ quasi-upper triangular. The diagonal of $\mathbf{U}$ contains blocks that directly encode the eigenvalues $\lambda_j = \sigma_j + i \omega_j$, such that clipping can be realized as $\tilde{\sigma}_j = \text{min}(\sigma_j, -\epsilon)$, leading to a stabilized $\tilde{\mathbf{\Theta}}_{lin} = \mathbf{Q} \tilde{\mathbf{U}} \mathbf{Q}^\top$. While such a clipping makes the linear core dense, it is beneficial compared to e.g. a linear damping, which would also affect otherwise stable modes equally, resulting in an overdamped system.

As explained in Section~\ref{sec:dim_red}, the EEG signal was reduced to two channels ($\text{CSP}_1$, $\text{CSP}_4$) which have a strong contrast in narrowband power under the two DBS conditions. It is hence easy to assume that DBS will lead to different dynamical regimes, which we exploit by training an E-SINDy for each DBS condition separately, resulting in two models $\mathbf{\Theta}^{\text{ON}}$ and $\mathbf{\Theta}^{\text{OFF}}$. Note that we refer to the models \textit{pars-pro-toto} by the parameter matrices ($\mathbf{\Theta}^{\text{ON}}$, $\mathbf{\Theta}^{\text{OFF}}$), as the same fixed library is used for both.
Each of the models is fitted on DBS \ON or \OFF data exclusively. For this purpose, each 6\,s trial is split at the t=4\,s point into training (0--4\,s) and testing (4--6\,s) data. Splitting within individual trials is an overfitting on temporal non-stationarity. Here, however, the aim is to use the models for generative purposes and not for predicting unseen data, hence we deem training on a richer training distribution acceptable. Splitting to 2\,s of testing data is a trade-off between containing the actual past (4\,s) within the training sample, while providing enough samples for model differences to manifest during forecasting and providing sufficient data to estimate forecast accuracy. Before fitting E-SINDy, the time series of two channels, and more importantly, their derivative time series, are normalized (z-scaled, with mean and standard deviation calculated across all data).

\subsubsection{Selecting hyperparameters for modeling}
Selecting the maximum derivative order, $k-1$, requires balancing model complexity against noise amplification. While minimizing $k$ is desirable to avoid the noise inflation inherent to higher-order numerical differentiation, the system must remain sufficiently expressive to capture the temporal dynamics of the signal. Analytically, the linear core of a second-order system ($k=2$, including the first derivative in the state vector) is restricted to a single harmonic oscillation. A fourth-order system expands this to allow for a single sideband modulation of a dominant carrier frequency. We selected a sixth order system because it is the minimum order required to admit a dominant carrier, with independent upper and lower side-bands. Two sidebands are required to create asymmetric modulation of the envelope which is clearly visible in our data (see Results section, Figure~\ref{fig:forecasting_results} for examples) and resembles data reported on beta bursts of the cortico-basal-ganglia network~\cite{Abdi-Sargezeh2025, Tinkhauser2017b}. Differentiation was achieved by applying a smoothing sinc kernel over a modified Gaussian window (degree=6, half-width=10)~\cite{Schmid2022}, a more robust alternative to the otherwise common Savitzky-Golay filters.

For the library $\Lambda(\mathbf{X})$, we chose to limit the candidates to second order terms, i.e., $\Lambda(\mathbf{X})= \left[1,\,\mathbf{X},\,\mathbf{X}^{\text{P2}}\right]$. This limits complexity while allowing for the most simple nonlinear interaction.
Selecting the threshold for the STLSQ ($\eta$) and the eigenvalue clipping parameter ($\epsilon$) is done by a grid search over $\eta\in\{0.3,0.4,0.5,0.6,1.0,1.1,\ldots,1.5,1.8,2\}$ and $\epsilon\in\{0,10^{-5},10^{-3},0.1,0.5,0.8\}$ and heuristically selecting a trade-off between forecast accuracy and DBS contrast, which are explained in the following section.

\subsection{Forecasting spatially filtered EEG data}
Given a model $\mathbf{\Theta}$ and an initial value $\mathbf{x}_0\in\mathbb{R}^{C\cdot k}$, numeric integration can be used to produce a forecast up to a forecast horizon $\tau$. We use a Runge-Kutta method (RK4)~\cite{Butcher2016} with fixed step size (equivalent to the sampling rate of the signal) for computational efficiency to produce forecasts $\text{FC}(\mathbf{x}_0, \mathbf{\Theta}, \tau)\in\mathbb{R}^{\tau\times C\cdot k}$.

With two models fitted and two regimes (DBS \ON and \OFF), forecasting (FC) for a time horizon of $\tau$ leads to four different forecast types: Two \textit{factual} forecast types in which the initial value, $\mathbf{x}_0^{\text{OFF}}$ (from a DBS \OFF trial) or $\mathbf{x}_0^{\text{ON}}$ (from a DBS \ON trial), matches the model
$\text{FC}(\mathbf{x}_0^{\text{ON}}, \mathbf{\Theta}^{\text{ON}}, \tau)$,
$\text{FC}( \mathbf{x}_0^{\text{OFF}}, \mathbf{\Theta}^{\text{OFF}}, \tau)$,
and two \textit{counterfactual} forecast types,
$\text{FC}(\mathbf{x}_0^{\text{ON}}, \mathbf{\Theta}^{\text{OFF}}, \tau)$,
$\text{FC}( \mathbf{x}_0^{\text{OFF}}, \mathbf{\Theta}^{\text{ON}}, \tau)$.

It is common to look at the bandpower of narrowband signals for aDBS control, or general BCIs, in a continuous motor context. Hence we are here interested in the ability to forecast the future signals power, and not in the phase component. Over the forecasting horizon we therefore compared the Hilbert envelope of the true data $\text{Env}(\mathbf{X})\in \mathbb{R}^{\tau \times C}$ with the envelope of the forecasts $\text{Env}(\hat{\mathbf{X}})$, $\hat{\mathbf{X}}:=\text{FC}(\cdot, \cdot, \tau)$. We quantify the difference as the root mean squared error of the envelope signals,
\begin{equation}
\text{eRMSE} = \sqrt{ \frac{1}{\tau C} \sum_{c=1}^{C} \sum_{t=1}^{\tau} \Big( \text{Env}(\mathbf{X})_{c,t} - \text{Env}(\hat{\mathbf{X}})_{c,t} \Big)^2 }\quad,
\end{equation}
which is a renormalized Frobenius norm of the signal differences, $\text{eRMSE} = \frac{1}{\sqrt{C \tau}} \big\| \text{Env}(\mathbf{X}) - \text{Env}(\hat{\mathbf{X}}) \big\|_F$.

For quantifying the forecast accuracy via eRMSE we consider the factual forecasts ($\text{FC}(\mathbf{x}_0^{\text{ON}}, \mathbf{\Theta}^{\text{ON}}, \tau)$,
$\text{FC}( \mathbf{x}_0^{\text{OFF}}, \mathbf{\Theta}^{\text{OFF}}, \tau)$). If the models are useful in capturing domain specific dynamics, we further expect the counterfactual to have a higher eRMSE. If both models would agree on all forecasts, they could not be used for a \textit{what-if} simulation, which is the purpose of building the models in the first place.

As a sanity check that the identified models capture the time domain dynamics of the data, rather than reproducing envelope statistics over the forecast horizon, we evaluated one-step-ahead predictions. For each trial, the factual model was integrated for a single step ($\Delta t = 1/300\,\text{s}$) from the initial value at $t=4\,\text{s}$. The absolute error between the predicted and the observed next sample of $\text{CSP}_1$ and $\text{CSP}_4$ was compared against a naive steady-state prediction ($\hat{\mathbf{x}}_{t+\Delta t}=\mathbf{x}_t$). The errors of both predictions were compared per channel and DBS condition on the same trials with a Wilcoxon signed-rank test.

To capture whether a model produces a forecast resembling the properties of the regime it was trained on, we again use a regularized LDA model for the selected two CSP channels (see section~\ref{sec:dim_red}). In particular, the decision score ($s$) of the LDA is used to define the recovery as follows.

For each trial $i$ two forecasts are generated from the initial value at t=4\,s ($\mathbf{x}_{t=4}^{i}$): $\text{FC}_{\text{OFF}}^{i}=\text{FC}(\mathbf{x}_{t=4}^{i}, \mathbf{\Theta}^{\text{OFF}}, \tau=2)$ and $\text{FC}_{\text{ON}}^{i}=\text{FC}(\mathbf{x}_{t=4}^{i}, \mathbf{\Theta}^{\text{ON}}, \tau=2)$. We write $\mathcal{F}_{\mathbf{\Theta}^c}$ for the set of forecasts generated by the model of regime $c \in \{\text{ON}, \text{OFF}\}$, and $f_i \in \mathbb{R}^{2}$ for the log-bandpower features of the forecasts $\text{CSP}_1$ and $\text{CSP}_4$ components of $\text{FC}_{c}^{i}$ (note that the forecast is integrating the full state-space, including the derivatives of $\text{CSP}_1$ and $\text{CSP}_4$. We ignored the derivatives for the bandpower calculation). The LDA classifier assigns a label $\hat{y}_i$ based on the decision score $s(f_i)$ as
\begin{equation}
\hat{y}_i =
\begin{cases}
  \text{ON}  & \text{if } s(f_i) > 0,\\
  \text{OFF} & \text{otherwise.}
\end{cases}
\end{equation}
The recovery $R_{c}$ is then defined as the rate of forecasts which match the regime the generating model was trained on
\begin{equation}
R_{c} = \frac{1}{|\mathcal{F}_{\mathbf{\Theta}^c}|}
        \sum_{i \in \mathcal{F}_{\mathbf{\Theta}^c}} \indicator\left[\hat{y}_i = c\right].
\end{equation}
Both models should be able to recover their regime for our simulation purposes, hence we quantify the balanced recovery as $R_{\text{bal}} = \tfrac{1}{2}\left(R_{\textsc{on}} + R_{\textsc{off}}\right)$.

\subsection{Training control strategies based on model forecasts}

In this section, we describe how the two models ($\mathbf{\Theta}^{\text{ON}}$, $\mathbf{\Theta}^{\text{OFF}}$) are used as a surrogate brain to train control algorithms for aDBS. With access to brain signals only, the objective for aDBS would be to keep the brain in a beneficial state most efficiently.
In this context, efficiency is energy efficiency, since we are not able to model other DBS parameters, given that the data used to derive the models contained only one DBS configuration. We first introduce how forecasting with the models can be used to simulate how the brain reacts to different DBS intensities (\textit{simulation environment}), and afterwards introduce the set of specific controllers that we applied within the simulation environment.

\subsubsection{The simulation environment}
Assuming that DBS is having a positive effect, as is the whole purpose of why it is used therapeutically, we consider DBS \ON signals to reflect the beneficial state. Hence, we can reuse the LDA classifier described in the context of the recovery metric, to define beneficial states ($s(f_j) > 0$) and non-beneficial ($s(f_j) < 0$) for a given forecast period $j$.

Now we use model forecasts as a simulation environment in which a controller defines the stimulation intensity $a$ (control action). The data contain only DBS \OFF and DBS \ON recordings, which we set to $a=0$, and $a=1$ respectively. As a first approximation, we use a linear mixture of both models
\begin{equation}
\mathbf{\Theta}^a = \mathbf{\Theta}^{\text{OFF}} + a(\mathbf{\Theta}^{\text{ON}} - \mathbf{\Theta}^{\text{OFF}}),
\end{equation}
with $a\in\left[0,1\right]$, to simulate the brains reaction to $a$, by forecasting with $\text{FC}(\mathbf{x}_0,\mathbf{\Theta}^a,\tau)$.
While the initial value $\mathbf{x}_0$ could be any vector in $\mathbb{R}^{C\cdot k}$, we chose to take samples from the existing data. This has two advantages of first ensuring that the initial value is from a realistic distribution, and second, a label about the DBS condition is known ($\mathbf{x}_0^{c}$, with $c\in\{\text{ON}, \text{OFF}\}$).

The mixture model and initial conditions are then turned into an episodic reinforcement learning (RL) problem. We chose a period length of 2\,s as a trade-off between realism and providing sufficient time for control actions to impact the dynamics. The longer we simulate the further the forecast will deviate from the realistic initial values. On the other hand, a controller would need sufficient time for a control action to manifest in the signal. Looking at how the two channel bandpower estimates evolve during forecasting (using a 500\,ms rolling window), the largest changes take place up to 2\,s into the future (see appendix Figure~\ref{xfig:500ms_rolling_window_estimation}). The 2\,s horizon is hence a practical choice and aligns with the horizon used for selecting the model parameters. A 4\,s horizon led to qualitatively the same controller results at a higher computational cost.

A controllers performance is evaluated based on the signal of this 2\,s forecast. Let $\hat{\mathbf{X}}_i\in\mathbb{R}^{T\times C}$ be the forecasted data for an initial value $\mathbf{x}_0^{i}$, $f_i$ the log-bandpower of the two CSP channels from this forecast, and $s(\cdot)$ the decision score of the LDA classifier (see schema in Figure~\ref{fig:approach_detail_flowchart}). The performance is calculated as $r_i = \text{R}(s(f_i)) - \text{Cost}(a)$, with the reward

\begin{equation}
\text{R}(s(f_i)) =
\begin{cases}
  +2  & \text{if } s(f_i) > 0,\\
  -2 & \text{otherwise,}
\end{cases}
\end{equation}
and a cost of the applied intensity $\text{Cost}(a)=c\int a\,\text{dt}$, for which we calibrate $c=1$ such that keeping $a=1$ over a 2\,s period (the cDBS equivalent) yields a reward of zero.
To truly have an adaptive response, we allow the controller to change the intensity in decision intervals of $\Delta t=250\,\text{ms}$ instead of just once at forecast onset. To make an informed decision, the controller is provided the bandpower estimates ($\tilde{f}_t$) of the preceding $\tau_{b}=2\Delta t=500\,\text{ms}$ interval and the decision score (s($\tilde{f}_t$)) at every decision point $t$. The $500\,\text{ms}$ window was chosen to be similar to feature estimation windows used in previous aDBS studies~\cite{Little2013, Stanslaski2024}. Furthermore, we observe that the bandpower estimation is stabilizing with respect to the LDA classifier performance at about $500\,\text{ms}$ (see appendix Figure~\ref{xfig:bandpower_window_sizes_vs_balanced_acc}).
Even with bounded $\text{a}\in\left[0, 1\right]$ the forecasts could pass a divergence threshold, which we set at $50\times$ the maximum training-set amplitude. If the threshold is exceeded, the forecasting and evaluation is terminated and a penalty of $R_{\text{blow}}=-10$ is accounted for.
The simulation loop for a single episode is provided in Algorithm~\ref{alg:simulation_loop}. Although a finite horizon of $\tau=2\,\text{s}$ is used in the simulation environment, the controller never receives information about the finite horizon or current distance to the horizon. This is important to prevent optimization targeting for this finite horizon (e.g., always turn DBS off at decision intervals close to the end, since the change in dynamics takes time to manifest, while the energy cost would be paid immediately).

\begin{algorithm}[H]
\caption{Forecast-based aDBS control environment (one episode)}
\label{alg:simulation_loop}
\begin{algorithmic}[1]
\Require models $\mathbf{\Theta}^{\text{OFF}}, \mathbf{\Theta}^{\text{ON}}$;
         classifier $s(\cdot)$; controller $\pi$; horizon $\tau$; decision interval $\Delta t$; seed sample index j;
\State $\mathbf{H} \gets$ the $k$ samples up to and including $j$
       \Comment{realistic initial value and history}
\State $\mathbf{x}_0 \gets \mathbf{H}[-1]$;\quad buffer $\gets \mathbf{H}$;\quad $E \gets 0$
    \Comment{initial value, ring buffer, and total energy}
\State forecasts $\gets$ empty list
\State $obs \gets s(\tilde{f}_t)\cup\tilde{f}_t$ \Comment $\tilde{f}_t$ bandpower computed over the buffer
\For{$n = 1, \dots, N$ \textbf{where} $N = \tau/\Delta t$}
  \State $a_n \gets \pi(obs)$               \Comment{controller $\pi$ defines the stimulation energy}
  \State $\mathbf{\Theta}^{a} \gets \mathbf{\Theta}^{\text{OFF}} + a_n\!\left(\mathbf{\Theta}^{\text{ON}} - \mathbf{\Theta}^{\text{OFF}}\right)$
  \State $\hat{\mathbf{X}} \gets \text{FC}(\mathbf{x}_0, \mathbf{\Theta}^{a}, \Delta t)$
                                            \Comment{RK4 integration up to $\Delta t$}
  \If{$\lVert\hat{\mathbf{X}} \rVert_\infty >$ divergence threshold}
    \State \Return $R_{\text{blow}}$        \Comment{forecast diverged, early stop}
  \EndIf
  \State buffer.push\_samples($\hat{\mathbf{X}}$)
  \State forecasts.append($\hat{\mathbf{X}}$)
  \State $obs \gets s(\tilde{f}_t)\cup\tilde{f}_t$  \Comment $\tilde{f}_t$ bandpower computed over the buffer
  \State $E \gets E + a_n \Delta t$                 \Comment{accumulate energy spent in the $\Delta t$ interval}
  \State $\mathbf{x}_0 \gets\hat{\mathbf{X}}\left[-1\right]$             \Comment{last value of forecast is new initial value}

\EndFor
\State $f_j \gets$ bandpower from concatenated forecasts  \Comment{$f_t$ bandpower of the full window $\tau$}
\State \Return $R(s(f_j)) - \text{Cost}(E)$
\end{algorithmic}
\end{algorithm}

\subsubsection{Control algorithms}
We selected the following control strategies to be evaluated within our simulation environment.

\begin{itemize}
    \item \textit{Single threshold control} (threshold): Single threshold based control represents the most simple \textit{bang-bang} type of control and was the first control applied in aDBS for PD~\cite{Little2013}. The controller turns DBS ON ($a=1$) or OFF ($a=0$) depending on whether a physiomarker crosses a predefined threshold in an upward or downward direction. Threshold based control is available on the Percept\texttrademark{} PC IPG (Medtronic, Minneapolis, MN, USA) and is still being actively researched~\cite{Busch2024}. The controllers action at decision point $k$ is the indicator function with respect to a threshold ($s^{*}$):
    \begin{equation}
    a_k = \indicator_{\{s_k > s^{*}\}}\,;
    \label{eq:threshold_action}
    \end{equation}

    \item \textit{Proportional control} (proportional): Proportional control is the other type of aDBS control available on commercial IPGs. It scales the DBS amplitude proportional to a physiomarker's deviation from a threshold ($a\in\left[0, a_{max}\right]$~\cite{Rosa2015, Bocci2021, Arlotti2021}. It is available in the AlphaDBS\textregistered{} device (Newronika S.p.A., Milan, Italy). With proportional controller, the action $a_k$ at decision point $k$, given a threshold ($s^{*}$) and proportionality gain ($k_p$) is given by:
    \begin{equation}
    a_k = \operatorname{clip}\!\Big(\tfrac{1}{2} + k_p\,(s^{*} - s_k),\; 0,\; 1\Big)
    \label{eq:prop_action}
    \end{equation}

    \item \textit{Proportional-Integral-Derivative control} (PID): PID control adjusts DBS according to the deviation of a physiomarker from a threshold, but not only proportional (P), but also relative to the accumulated historical deviation (I), and the rate of change of the deviation (D).
    \begin{align}
    u_k &= \tfrac{1}{2} + k_p\left(e_k + \frac{1}{T_i}\sum_{j\le k} e_j\cdot\Delta t + T_d\cdot d_k\right), \quad &e_k = s^{*} - s_k,\\
    d_k &= d_{k-1} + \alpha\left( \frac{e_k - e_{k-1}}{\Delta t} - d_{k-1} \right),
         \quad &\alpha = \frac{\Delta t}{\tau_d + \Delta t}, \label{eq:pid-deriv} \\
    a_k &= \operatorname{clip}(u_k, 0, 1)
    \label{eq:pid_action}
    \end{align}

    PI(D) has previously been investigated in bottom-up simulation~\cite{Fleming2020, Kuzmina2025} and with a small cohort of patient participants in the PI version~\cite{Schmidt2023}. Note, both the \textit{proportional} and the \textit{threshold} controller are special cases of PID, which can be realized by setting $T_d = 0$, $T_i \xrightarrow{}\infty$ and $k_p=0$ respectively.

    \item \textit{Proximal Policy Optimization} (PPO): PPO is a policy gradient RL algorithm serving as a modern, but data-hungry deep-learning baseline. It should be able to learn complex, nonlinear DBS strategies including long-term optimization within the forecasted environment, without relying on predefined clinical heuristics (thresholds). PPO was selected as one representative of the RL for being a stable, continuous-action, on-policy model which was previously used within bottom-up simulation studies~\cite{Krylov2020, Kuzmina2025}.

\end{itemize}

\subsubsection{Data split}
The simulation environment in Algorithm~\ref{alg:simulation_loop} requires 500\,ms of data for the initial buffer filling and bandpower estimation. Hence, every sample in each trial with $t\geq500\,\text{ms}$ can be used as a seed for the algorithm. These samples of course are highly correlated, hence we subsample uniformly to 1000 samples per DBS condition for the training set (2000 samples in total), drawn from $t\in[0, 4]$ to match the training/test split of the E-SINDy fitting, and another 1000 samples per DBS condition for the testing set, drawn from $t\in(4, 6]\,s$.

The threshold, proportional, and PID controllers are optimized using Bayesian optimization on the training set with 50 random starts and 50 guided searches. Parameter ranges are provided in Table~\ref{tab:controller-search-spaces}. The optimization is implemented using the optuna~\cite{Akiba2019optuna} Python library.

The PPO controller is using the Stable-Baselines3~\cite{stable-baselines3} implementation within a gymnasium environment~\cite{towers2024gymnasium}. The RL environment and PPO parameters are provided in Table~\ref{tab:ppo-hyperparameters}. Gymnasium is also used for the final evaluation of all trained control strategies.

\begin{table}[t]
\centering
\small
\renewcommand{\arraystretch}{1.25}
\begin{tabular}{@{}llllp{0.44\linewidth}@{}}
\toprule
\textbf{Controller} & \textbf{Parameter} & \textbf{Range} & \textbf{Scale} & \textbf{Rationale for the bounds} \\
\midrule
threshold
& $s^{*}$ & $[-12.27,\, 9.25]$ & lin
& \textit{LDA decision score threshold} - Measured per run as the min/max observed over all-ON and all-OFF
  rollouts (300 episodes). \\
\midrule
\multirow{2}{*}{proportional}
& $s^{*}$ & $[-12.27,\, 9.25]$ & lin
& As above; the same measured score support. \\
& $k_p$ & $[10^{-2},\, 10^{1}]$ & log
& \textit{Proportionality gain} - The graded region spans $1/k_p$ score units: at $k_p=0.01$ the action never leaves $[0.43, 0.67]$ ($\approx$ cDBS with a=0.5); at $k_p=10$ the policy differs from binary bang-bang on $0.4\%$ of the support only.\\
\midrule
\multirow{5}{*}{PID}
& $s^{*}$ & $[-12.27,\, 9.25]$ & lin
& \textit{LDA decision score setpoint} - As above, the min/max of $s$ over
all-ON and all-OFF rollouts. \\
& $k_p$ & $[10^{-2},\, 10^{1}]$ & log
& \textit{Proportionality gain} - As above; $k_p$ is the same quantity in both
controllers, so the two rungs must admit the same gains. \\
& $T_i$ & $[0.05,\, 16]$\,s & log
& \textit{Integral time} - equivalent to $0.2$ to $64$ decision steps: below one control interval at
the lower end, $8\times$ the episode at the upper, beyond which the integral is a
constant offset and indistinguishable from $T_i\to\infty$. \\
& $T_d$ & $[10^{-3},\, 2]$\,s & log
& \textit{Derivative weight} - $0.004$ steps is indistinguishable from a pure PI
controller; $8\times$ decision step is where the derivative term dominates the proportional one. \\
& $\tau_d$ & $[0,\, 0.5]$\,s & lin
& \textit{Derivative low-pass} - Sets the differencing horizon; linear because $0$ (the raw one-step difference) is meaningful and
reachable, and the upper bound matches the $0.5$\,s lookback $s$ is averaged over. \\
\bottomrule
\end{tabular}
\vspace{1em}
\caption[Controller search spaces]{%
  Search spaces for the three parametric controllers. Each was optimised with
  Optuna's TPE sampler under an identical budget of $100$ trials ($50$
  quasi-random start-up, $50$ guided), maximising the paired per-episode
  advantage over continuous stimulation on $2000$ tuning episodes, evaluated
  on $2000$ disjoint validation episodes, and reported on the $2000$ held-out
  test episodes.}
\label{tab:controller-search-spaces}
\end{table}

\begin{table}[t]
\centering
\small
\renewcommand{\arraystretch}{1.25}
\begin{tabular}{@{}llp{0.52\linewidth}@{}}
\toprule
\textbf{Group} & \textbf{Setting} & \textbf{Value and rationale} \\
\midrule

\multirow{4}{*}{Policy}
  & Architecture & \texttt{MlpPolicy}, \textsc{stable-baselines3} default, with two hidden layers of $64$ units, separate policy and value heads. \\
  & Observation & $3$-dimensional: the decoder score $s_k$ of the lookback
    window followed by the two \textsc{csp} channel powers. The score is the
    first element, which is also the only input the \textit{threshold}, \textit{proportional} and PID receive.\\
  & Action & Scalar $a_k \in [0,1]$, applied as a constant amplitude over the
    decision interval ($\Delta t = 250\,\text{ms}$). \\
  & Exploration & gSDE (\texttt{use\_sde}), resampled every $4$ steps,
    $\log \sigma_0 = -1$. State-dependent rather than per-step white noise
    because the action is held constant across a $250\,\text{ms}$ decision. \\
\midrule

\multirow{5}{*}{Optimisation}
  & Learning rate & $3 \times 10^{-4}$ (Adam, \textsc{sb3} default). \\
  & Clip range & $0.2$. \\
  & Entropy / value coef. & $0.01$ / $0.5$. \\
  & Epochs, minibatch & $10$ epochs per update, minibatch $256$. \\
  & Gradient clipping & Max-norm $0.5$. \\
\midrule

\multirow{4}{*}{Returns}
  & $\gamma$ & $1.0$ (undiscounted). An episode is a fixed $8$ decisions and
    the reward is defined per episode---no discounting. \\
  & $\lambda_{\mathrm{GAE}}$ & $1.0$ for the terminal objective. The reward is
    a single terminal payout plus per-step energy cost. No signal for a shorter-horizon advantage estimate to exploit. \\
  & Episode length & $8$ decisions at $\Delta t = 250\,\text{ms}$ ($2$\,s final horizon). \\
  & Divergence & $-10$ for blow up > 50$\times$ max. observed amplitude. \\
\midrule

\multirow{4}{*}{Rollout}
  & Parallel episodes & $64$ lockstep episodes, balanced $32/32$ over DBS \ON and \OFF. \\
  & Steps per update & $32$ per environment, i.e.\ $64 \times 32 = 2048$
    transitions per policy update ($48$ updates per seed). \\
  & Batch resampling & $64$ balanced redraws at every new
    episode---total of $12{,}500$ episodes per seed over the training set. \\
  & Budget & $100{,}000$ timesteps per seed, $5$ seeds
    ($0$--$4$). Held-out curve flat from ${\sim}65{,}000$ onward in visual inspection. Results are reported for the seed with the median test performance (seed $1$); paired advantage over cDBS across seeds: $0.222 \pm 0.018$ (mean $\pm$ SD). \\
\bottomrule
\end{tabular}
\vspace{1em}
\caption[PPO hyperparameters]{%
  Hyperparameters of the \textsc{ppo} policy.}
\label{tab:ppo-hyperparameters}
\end{table}

\subsection{Statistics}
The used tests are two-sided and non-parametric, as log-bandpowers, envelope errors and reward differences are not normally distributed. Rewards take discrete values in the proposed setting. Tests were computed with SciPy~\cite{Scipy2020}. For multiple testing, $P$-values were Bonferroni-corrected by multiplying with the number of comparisons and significance is reported against the corrected values ($\text{P}_{BF}$, $\alpha=0.05$).
Statistical tests are used at the following results:

\textit{Forecast accuracy (Figure~\ref{fig:forecasting_results}~D, J).} For each model, the eRMSE of factual and counterfactual forecasts was compared per channel with a Mann--Whitney U test, as both groups consist of independent trials. Correction was applied over the two channels per model.

\textit{Forecast bandpower (Figure~\ref{fig:forecast_classification}~F, G).} For each model, the log-bandpower of forecasts starting from $\mathbf{x}_0^{\text{OFF}}$ and $\mathbf{x}_0^{\text{ON}}$ was compared per channel with a Mann--Whitney U test (72 vs.\ 72 trials), corrected over the two channels per model.

\textit{One-step-ahead prediction (appendix Figure~\ref{xfig:one_step_forecasts}).} The absolute one-step prediction error of the factual model and that of a naive steady-state prediction were obtained on the same trials and compared with a Wilcoxon signed-rank test ($n=72$ trials per channel and DBS condition), corrected over the four channel--condition combinations.

\textit{Controller comparison (Figure~\ref{fig:controller_performance}~A).} Each controller and cDBS were evaluated on identical initial conditions, which yields a paired reward difference per test episode. The 2000 test episodes originate from samples within 144 trials and are therefore not independent. As a consequence, we averaged the paired differences within each trial and tested the 144 trial means against zero with a Wilcoxon signed-rank test, corrected over the four aDBS controllers. Comparisons (PID vs.\ PPO, \textit{threshold} vs.\ \textit{proportional}) of net reward and mean action (energy) used the same trial-level paired Wilcoxon signed-rank test, corrected over these four comparisons. PPO results are reported for the seed with the median test performance; the variability across the five training seeds is reported as mean $\pm$ SD.

\subsection{Code and data availability}
The data are publicly available via the Radboud Data Repository~\cite{dold2026dataset}. The code for model fitting, simulation and controller optimization will be made publicly available at \url{https://github.com/matthiasdold/dbs_sim} upon publication of the peer-reviewed version of this article, and the version used for this article will additionally be archived on Zenodo with a persistent DOI. Until then, it is available from the corresponding author upon reasonable request.

\section{Results}
Fitting two separate E-SINDy models to the recorded EEG data, characteristics of the narrow band CSP spatially filtered signal could be reconstructed. The results comprise the grid search and heuristic model selection, the forecasting quality of the selected models, and finally the use of the models in a control simulation environment.

\subsection{Grid search and model selection}
To find suitable E-SINDy models for the two DBS conditions ($\mathbf{\Theta}^{\text{OFF}}$, $\mathbf{\Theta}^{\text{ON}}$) we used a grid search over two hyperparameters, which are the threshold for SINDy's STLSQ optimization, and the clipping parameter for stabilizing the linear core (clipping real part of the eigenvalues). The objective is to identify models which are both realistically describing the near future in a forecast scenario, and being expressive about the two different DBS conditions. The realism is linked to how accurately the models can predict the future, measured using the eRMSE. How well the two models describe different dynamics according to the DBS condition they were trained on (ON or OFF), is quantified by the balanced recovery.

\begin{figure}
    \centering
    \includegraphics[width=\linewidth]{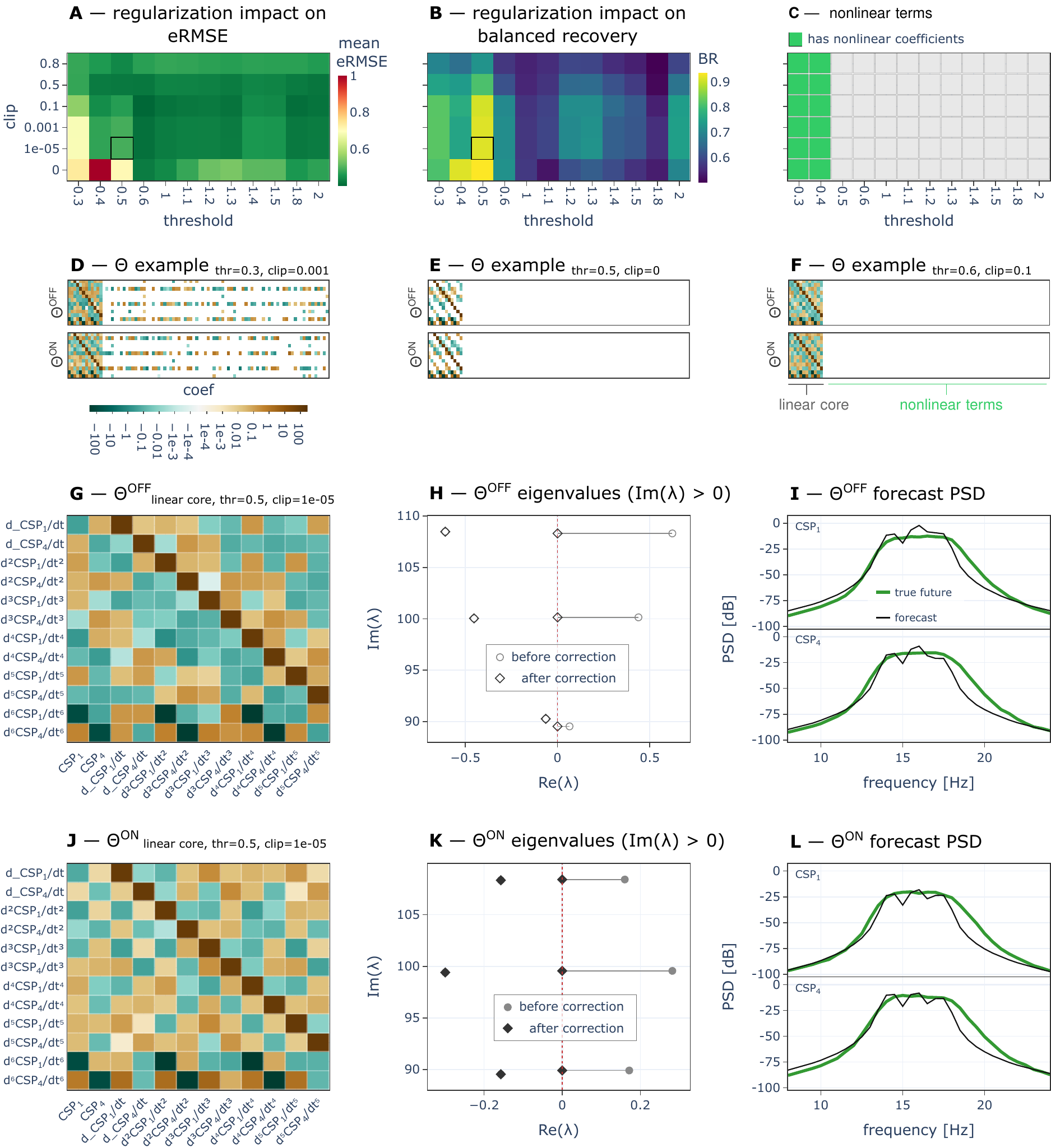}
    \caption{Parameter grid search and selected model. Panels \textbf{A-C} show the grid for different values of the \textit{threshold} parameter for the sequential thresholding least squares optimization and for the clipping parameters used to \textit{clip} the real part of the eigenvalues. Panel \textbf{A} shows the envelope root mean squared error (eRMSE) as measure of forecast accuracy. Panel \textbf{B} shows the balanced recovery as measure of how well both models, trained under DBS ON and OFF, recover characteristics of their training regime. The black frame in A and B marks the selected parameters ($\text{threshold}=0.5$, $\text{clip}=10^{-5}$).
    Panel \textbf{C} encodes the presence of nonlinear terms in the fitted models coefficient matrices. Green cells correspond to at least one of the models having nonlinear terms. Cells are gray otherwise. Panels \textbf{D-F} illustrate the model coefficient matrices ($\mathbf{\Theta}$) for example parameter values. Panels \textbf{G} and \textbf{J} show the linear core of the selected models for DBS OFF (G) and ON (J). All coefficient heatmaps share the same color coding (see D) using a logarithmic color scale. Panels \textbf{H} and \textbf{K} show the positive imaginary half plane of the eigenvalue spectrum and the impact of the clipping of eigenvalues $\lambda$ with $\text{Re}(\lambda) > -\epsilon$, the clipped values are shown close to the $\text{Re}(\lambda) = 0$ and are connected to their original positions. Panels \textbf{I} and \textbf{L} show the power spectral densities (PSD) of the forecast (black) and the actual observed future (green) for the two common spatial pattern channels ($\text{CSP}_1$, $\text{CSP}_4$). The carrier corresponding to the eigenvalues with $\text{Im}(\lambda) \approx 100$ is visible at $\approx16\,\text{Hz}$, with the two sidebands at approximately $14$ and $17\,\text{Hz}$.
    }
    \label{fig:threshold_clip_grid}
\end{figure}

The eRMSE (Figure~\ref{fig:threshold_clip_grid} A) over the two grid dimensions shows that applying clip values as small as $10^{-5}$ is improving accuracy over the non-clipped scenarios (clip=0). Increasing clipping value further has little effect on accuracy, until it reaches 0.8, at which point clipping over-damps the system and accuracy declines. The threshold parameter is optimal at 0.6, with acceptable accuracy as low as 0.4 provided that clipping with at least $10^{-5}$ is provided.
Balanced recovery (Figure~\ref{fig:threshold_clip_grid} B) is strongest for threshold values 0.5, followed by lower threshold. Using clipping becomes detrimental for balanced recovery a bit earlier as clip values of 0.5.

Higher threshold parameter make the models more sparse by eliminating smaller coefficients. This is having a structural impact in that for models with a threshold of 0.5 and greater do no longer contain any non-zero nonlinear coefficients (Figure~\ref{fig:threshold_clip_grid} C). Interestingly, the best balanced recovery is reached by the least regularized models, which do only contain linear coefficients. Figures~\ref{fig:threshold_clip_grid} D-F illustrate some coefficient matrices, showing the impact of different threshold and clipping parameters.
At threshold values lower than $0.3$, the fitted models started to diverge during forecasting the $2\,\text{s}$ horizon. Smaller values were therefore not considered.

We proceeded with models fitted with $\text{threshold}=0.5$ and $\text{clip}=10^{-5}$ as they provide the best balanced recovery ($R_{\text{bal}}=91.7\,\%$; $R_{\text{OFF}}=100\,\%$, $R_{\text{ON}}=83.3\,\%$) while having an acceptable eRMSE. We calculated the one-step ahead model prediction error against a naive steady state prediction as a sanity check to confirm that the models do in fact capture the system dynamics (see appendix Figure~\ref{xfig:one_step_forecasts}).
Both models predict significantly better than the naive prediction (Wilcoxon signed-rank test, $n=72$ trials per channel and DBS condition, $\text{P}_{BF}<10^{-12}$).

The coefficient matrices for both are shown in Figures~\ref{fig:threshold_clip_grid} G ($\mathbf{\Theta}^{\text{OFF}}$) and J ($\mathbf{\Theta}^{\text{ON}}$). The impact of clipping on the eigenvalue spectrum is illustrated in Figures~\ref{fig:threshold_clip_grid} H and K. Only eigenvalues ($\lambda$) with $\text{Im}(\lambda)>0$ are shown as the eigenvalues are always symmetric with respect to the $\text{Im}(\lambda)=0$ axis. Additionally, before clipping, the eigenvalues are close to being symmetric to the $\text{Re}(\lambda)=0$ axis as well (close to an energy conserving Hamiltonian system); the clipping moves the unstable ones onto $\text{Re}(\lambda)=-\epsilon$ (Figures~\ref{fig:threshold_clip_grid} H and K, appendix Table~\ref{xtab:eigenmodes}). Figures~\ref{fig:threshold_clip_grid} I and L show the power spectral densities (PSD) of the forecast and the actual observed future for the two CSP channels ($\text{CSP}_1$, $\text{CSP}_4$). The selected models, $\mathbf{\Theta}^{\text{OFF}}$ and $\mathbf{\Theta}^{\text{ON}}$, forecast the narrowband characteristics of both channels reasonably well. The models contain three modes with the central one at $\mathbf{\Theta}^{\text{OFF}}: \text{Im}(\lambda)=100 \rightarrow 100/2\pi=15.91\,\text{Hz}$ and $\mathbf{\Theta}^{\text{ON}}: \text{Im}(\lambda)=99.41 \rightarrow 99.41/2\pi=15.82\,\text{Hz}$ (Figure~\ref{fig:threshold_clip_grid} G - L), matching the central frequency of the band.

\subsection{Model base forecasting}
\begin{figure}
    \centering
    \includegraphics[width=\linewidth]{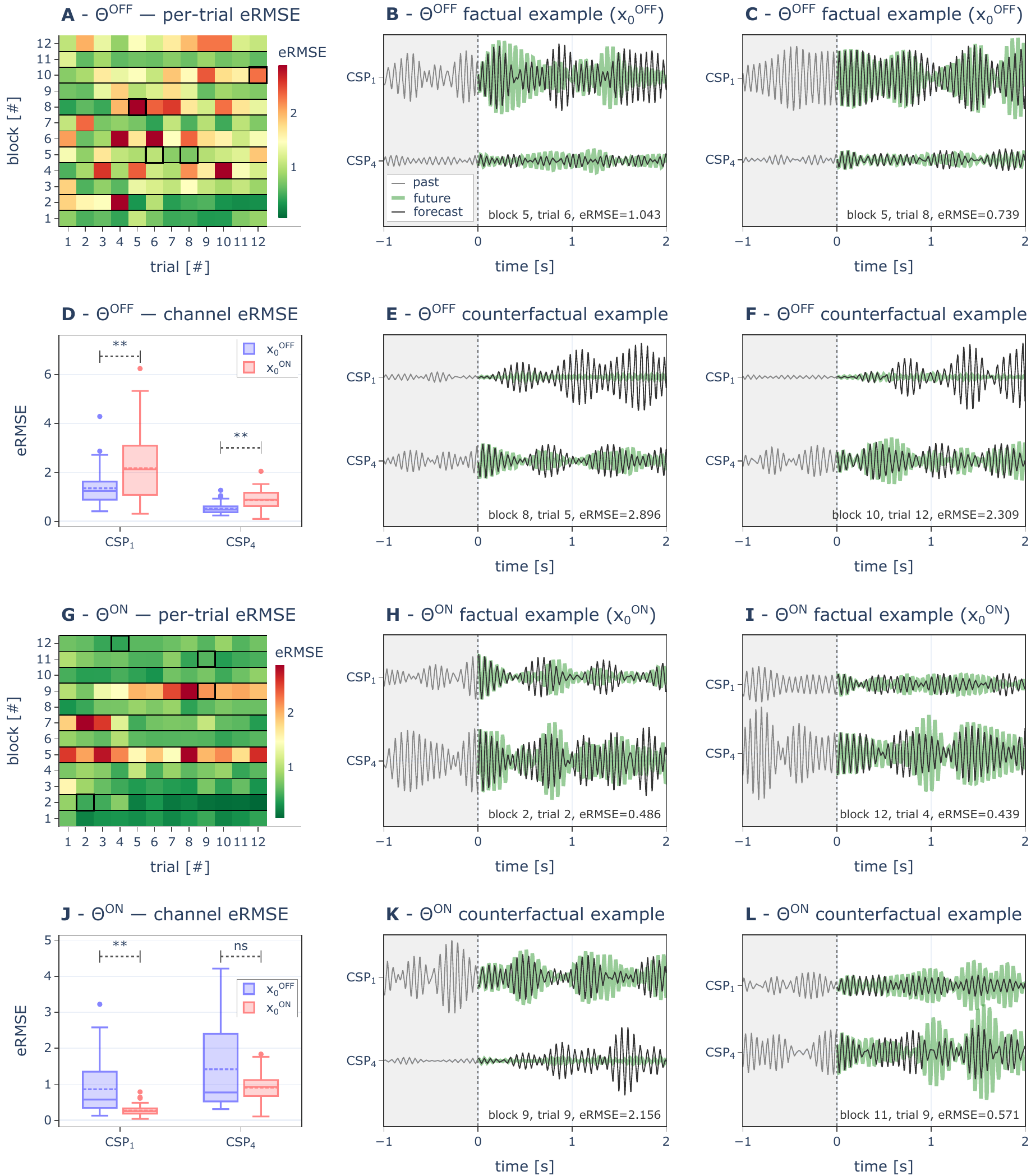}
    \caption{Forecasting examples and fit quality. The second and third columns (\textbf{B,C,E,F,H,I,K,L}) show factual and counterfactual forecasts of the two models with a forecast horizon of 2\,s. A \textit{factual} forecast corresponds to the initial value ($\mathbf{x}_0^{\text{c}}$) being from data of the same DBS condition $c\in\{\text{OFF}, \text{ON}\}$ the model $\mathbf{\Theta}^{c}$ was trained on. The forecast is \textit{counterfactual} otherwise. Shown are both CSP channels with the forecasted data in black, the past in gray (with gray background), and the observed future data in green. Panels \textbf{A} and \textbf{G} show the envelope root mean squared error (eRMSE) as mean across channels for the forecast of each of the 144 trials. Each trial with total duration 6\,s was split at t=4\,s with the eRMSE being calculated between the forecasts with initial value $\mathbf{x}_{t=4s}^{c}$ and the observed future for $t\in[4, 6]\,s$. The DBS condition was fixed during blocks of twelve trials each. Odd block numbers correspond to DBS OFF and even numbers to DBS ON. Black frames are drawn to the trials corresponding to the examples in the second and third column. Panels \textbf{D} and \textbf{J} show box plots of eRMSE per channel, colored according to the DBS condition of the initial value (DBS OFF in blue, DBS ON in red). Significance according to a two-sided Mann-Whitney U test (factual vs.\ counterfactual initial values, $n=72$ trials each) is indicated with (**) for $\text{P}_{BF}<0.01$ and (ns) for not significant (Bonferroni correction over the two channels).
    }
    \label{fig:forecasting_results}
\end{figure}

While actual precise forecasting is not strictly required for the models to be usable within a simulation environment, it is a link to realism of the simulated data. If the models are capable of accurately predicting the future, we assume they captured real dynamics present in the data. Figure~\ref{fig:forecasting_results} shows selected examples of factual (B-C for $\mathbf{\Theta}^{\text{OFF}}$, H-I for $\mathbf{\Theta}^{\text{ON}}$) and counterfactual forecasts (E-F for $\mathbf{\Theta}^{\text{OFF}}$, K-L for $\mathbf{\Theta}^{\text{ON}}$). The counterfactual examples show how the amplitude of at least one channel gets amplified above the observed range. Overall the counterfactual cases lead to significantly higher eRMSE in both channels for the $\mathbf{\Theta}^{\text{OFF}}$ model (Figure~\ref{fig:forecasting_results} D), and for the $\text{CSP}_1$ channel of the $\mathbf{\Theta}^{\text{ON}}$ model (Figure~\ref{fig:forecasting_results} J). A temporal structure in the eRMSE with blocks five, seven and nine (all DBS OFF) showing higher eRMSE compared to other DBS OFF blocks (odd numbers) is evident in Figure~\ref{fig:forecasting_results} G.

\subsection{Classifying model forecasts}
\begin{figure}
    \centering
    \includegraphics[width=\linewidth]{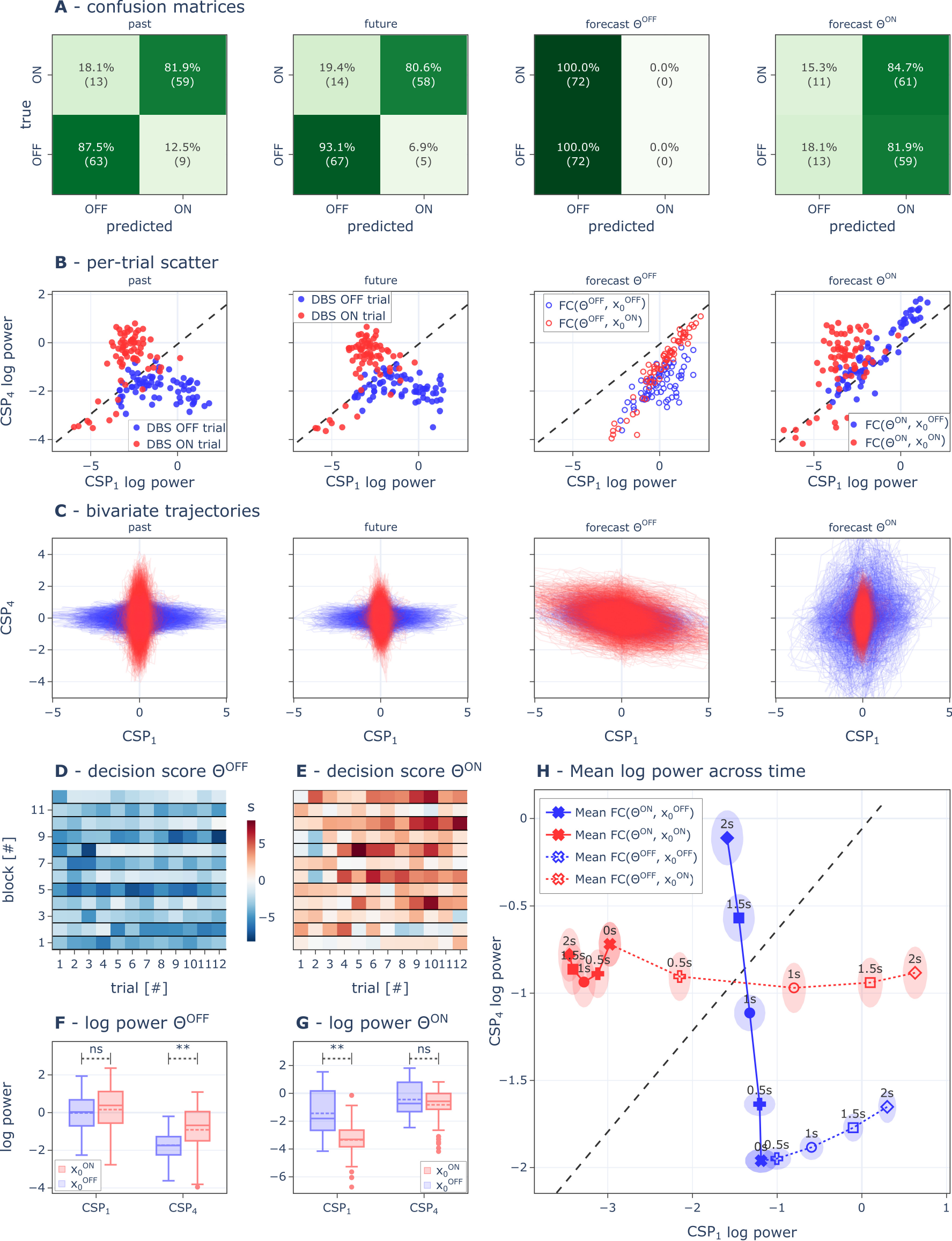}
    \caption{Classification of \textit{past}, \textit{future}, and \textit{forecast} data. \textit{Past} corresponds to all 144 trials (12 blocks of 12 trials each) real recorded data with trial time up to $t=4\,\text{s}$. It is the data that was used for fitting the E-SINDy models as well as the LDA classifier. \textit{Future} is the real recorded data for t>4\,s, up to the trial end at (6\,s). It was never used for fitting. \textit{Forecast} are available for each model ($\mathbf{\Theta}^{\text{OFF}}$, $\mathbf{\Theta}^{\text{ON}}$) where one forecast was created for each of the 144 trials. The initial value $\mathbf{x}_0^{\cdot}$ for the forecasts was always the last sample of each trials \textit{past} data ($t=4\,\text{s}$). A forecast was factual if the DBS condition of the initial value matches the DBS condition the forecasting model was trained on ($\mathbf{x}_0^{\text{OFF}}$ for $\mathbf{\Theta}^{\text{OFF}}$, and $\mathbf{x}_0^{\text{ON}}$ for $\mathbf{\Theta}^{\text{ON}}$). The forecast was counterfactual otherwise. Panel \textbf{A} shows the confusion matrices of classifying the from \textit{past}, \textit{future}, and \textit{forecasts}. Panel \textbf{B} shows scatter plots of the two channels log-bandpower for the same data groups. Red markers correspond to trial with DBS ON, blue markers to DBS OFF. Panel \textbf{C} shows bivariate trajectories with one trajectory per trial, colored as in B. Panels \textbf{D} and \textbf{E} show the LDA decision score (s) for the forecasts of $\mathbf{\Theta}^{\text{OFF}}$ and $\mathbf{\Theta}^{\text{ON}}$ respectively. Panels \textbf{F} and \textbf{G} show the distribution of log-bandpower for DBS OFF (blue) and DBS ON (red) per channel. Significance of a two-sided Mann-Whitney U test between the DBS conditions of the initial values ($n=72$ trials each) is marked with (**) for $\text{P}_{BF}<0.01$ (Bonferroni correction over the two channels). Panel \textbf{H} shows the mean log-bandpower for factual and counter factual forecasts for each model. Hollow markers correspond to $\mathbf{\Theta}^{\text{OFF}}$, filled markers to $\mathbf{\Theta}^{\text{ON}}$. The DBS condition of the initial value is encoded in color (blue for DBS OFF, red for DBS ON). The background ellipses are the standard error of mean.}
    \label{fig:forecast_classification}
\end{figure}

How well the model forecasts preserve the characteristics of the DBS condition they were trained on is quantified by using an LDA classifier trained on the same training data. Figure~\ref{fig:forecast_classification} A shows the confusion matrices for this classifier, evaluated in training sample (\textit{past}), and out of sample for real data (\textit{future}), forecasts from the $\mathbf{\Theta}^{\text{OFF}}$ model, and forecasts for the $\mathbf{\Theta}^{\text{ON}}$ model. The true label for the forecasts is the label of the initial value ($\mathbf{x}_0^{\text{OFF}}\rightarrow\textit{OFF}$, $\mathbf{x}_0^{\text{ON}}\rightarrow\textit{ON}$). The confusion matrices for \textit{past} and \textit{future} are qualitatively aligned, with higher accuracy for identifying the DBS OFF case. This is a first indication for the log-bandpower distributions (LDA features) to be similar in both data sets. An indication which is later confirmed by the plots in Figure~\ref{fig:forecast_classification} B and C. The confusion matrices for the forecasts provide additional detail about how the balanced recovery is composed: The models produce forecasts which look to the classifier as if they would originate from the models training data feature distribution. The $\mathbf{\Theta}^{\text{OFF}}$ is stronger in that sense, as it makes all forecasts look as if they are DBS OFF. The $\mathbf{\Theta}^{\text{ON}}$ model has a weaker impact with $84.7\,\%~(61)$ of forecasts starting from $\mathbf{x}_0^{\text{ON}}$ being classified as DBS ON. For forecasts starting from $\mathbf{x}_0^{\text{OFF}}$ still $81.9\,\%~(59)$ are classified as DBS ON. Accordingly, $15.3\,\%~(11)$ and $18.1\,\%~(13)$ forecasted trials are classified as DBS OFF.\\
Per trial scatter plots of the log-bandpower features for \textit{past}, \textit{future} and \textit{forecasts} are shown in Figure~\ref{fig:forecast_classification} B. A dashed line shows the binary decision boundary of the LDA model. The individual features show the same general alignment of the true \textit{past} and \textit{future} data. The \textit{forecasts} data shows a more linear alignment of the features with slightly higher overall power, especially for the $\mathbf{\Theta}^{\text{OFF}}$ model (initial values from both DBS conditions) and the $\mathbf{\Theta}^{\text{ON}}$ model for $\mathbf{x}_0^{\text{OFF}}$ initial values.\\
While Figure~\ref{fig:forecast_classification} B was showing aggregate scores over the full data window (log-bandpower), Figure~\ref{fig:forecast_classification} C is showing individual bivariate trajectories, overlaying 144 trajectories for each of \textit{past}, \textit{future} and \textit{forecasts}. While individual trajectories can hardly be identified, this plot is valuable to understand how stable the trajectories are over time. Both \textit{past}, and \textit{future} again align with slightly lower amplitudes for the \textit{future} data. The $\mathbf{\Theta}^{\text{OFF}}$ forecasts show the trajectories with inflated amplitudes for $\text{CSP}_1$, while the overall amplitude for $\text{CSP}_4$ is comparable with the \textit{future}. The $\mathbf{\Theta}^{\text{ON}}$ forecasts lead to a well located distribution with small $\text{CSP}_1$ and larger $\text{CSP}_4$ values for trajectories with $\mathbf{x}_0^{\text{ON}}$ (red, factual). Trajectories with $\mathbf{x}_0^{\text{OFF}}$ (blue, counterfactual), have their $\text{CSP}_4$ values increased, while $\text{CSP}_1$ is similar to the observed \textit{future} trajectories.\\
The decisions score (s) of the LDA classifier for forecasts from $\mathbf{\Theta}^{\text{OFF}}$ for each trial, including the block structure is show in Figure~\ref{fig:forecast_classification} D. The same for forecasts of $\mathbf{\Theta}^{\text{ON}}$ is shown in Figure~\ref{fig:forecast_classification} E. Again, the plots show that $\mathbf{\Theta}^{\text{OFF}}$ makes forecasts more clearly result in $\text{s}<0$, i.e., be classified as DBS OFF. For $\mathbf{\Theta}^{\text{ON}}$ we see a clearer difference between the factual and counterfactual cases, visible in the alternative row structure contrasts for block six and higher. Especially in the second half of the second block (data was recorded under DBS ON), the model forecasts do not achieve $\text{s}>0$, potentially linked to a slow non-stationarity which dominates the signal.\\
Figure~\ref{fig:forecast_classification} F and G show the log-bandpowers for each channel and DBS condition of the initial value for $\mathbf{\Theta}^{\text{OFF}}$ and $\mathbf{\Theta}^{\text{ON}}$ respectively. The difference of factual and counterfactual initial conditions manifests in different channels each. For $\mathbf{\Theta}^{\text{OFF}}$ counterfactual initial conditions ($\mathbf{x}_0^{\text{ON}}$) lead to a higher bandpower in $\text{CSP}_4$, while $\text{CSP}_1$ has higher bandpower in the counterfactual ($\mathbf{x}_0^{\text{OFF}}$) case for $\mathbf{\Theta}^{\text{ON}}$.\\
Looking at the bivariate bandpower, Figure~\ref{fig:forecast_classification} H shows how the mean and standard error of mean (SEM, colored background ellipses) evolve in a rolling bandpower evaluation over a 500\,ms window. At t=0\,s all data is real past data and hence does not depend on the forecast at all. The figure shows that mean bandpower is changed much stronger in the counterfactual cases (larger distance at 2\,s from the 0\,s start). The change is mostly confined to a single channel per model (see mostly horizontal and vertical change), which is in agreement with the significance results from Figure~\ref{fig:forecast_classification} F and G. The means of the counterfactual cases clearly cross the LDA decision boundary (dashed line), with approximately 1\,s and 1.5\,s of forecast required for the $\mathbf{\Theta}^{\text{OFF}}$ and $\mathbf{\Theta}^{\text{ON}}$ models respectively. The SEM is mostly constant under $\mathbf{\Theta}^{\text{ON}}$, while a clear increase is visible for $\mathbf{\Theta}^{\text{OFF}}$ in the counterfactual case. In the factual cases, $\mathbf{\Theta}^{\text{OFF}}$ is increasing the amplitudes for both channels, while $\mathbf{\Theta}^{\text{ON}}$ is mostly stale on $\text{CSP}_4$ and damping $\text{CSP}_1$.

\subsection{Control simulation environment}
The models $\mathbf{\Theta}^{\text{OFF}}$ and $\mathbf{\Theta}^{\text{ON}}$ are a condensed encoding of brain activity under DBS OFF and ON. We assume continuous change between both dynamical regimes is possible, and use the models to train control strategies by simulating the brain's response to a control action by means of forecasting.

\begin{figure}
    \centering
    \includegraphics[width=\linewidth]{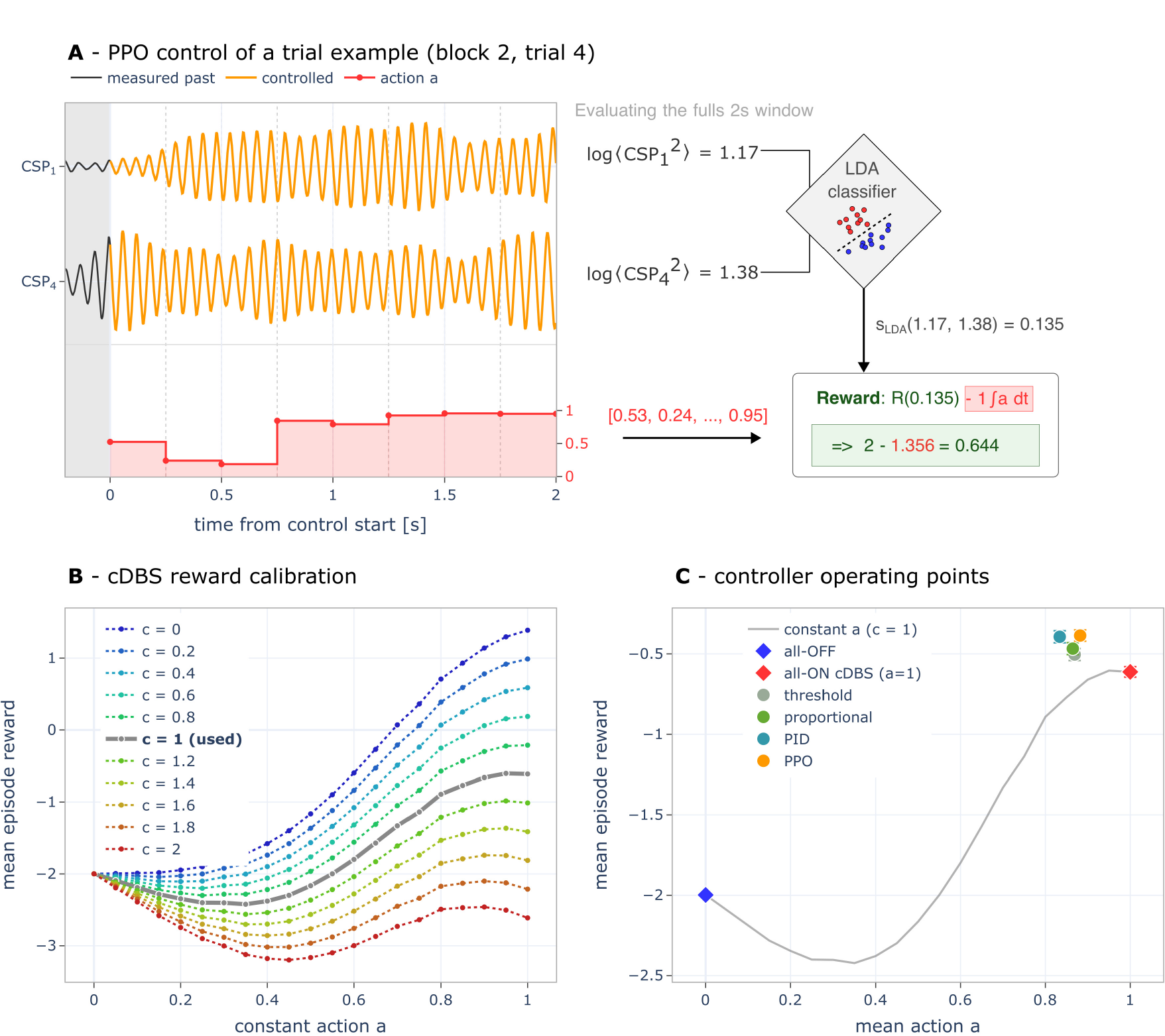}
    \caption{Control example and calibration. Panel \textbf{A} visualizes an aDBS with PPO in a single episode. Both CSP channels are shown with controlled/simulated signals (orange) and controller action (red) with markers at the according decision points ($\Delta t=250\,\text{ms}$, starting at $t=0\,\text{s}$). The full episode is classified and a reward calculated based on the classification and the average action. Panel \textbf{B} illustrates the calibration of the cost parameter by plotting the mean reward of episodes (y-axis) for different costs (c) assuming constant action control values (x-axis). Optimal choices with respect to the reward are marked with a star. The selected choice is $\text{c}=1$ (gray curve, bold legend label). Panel \textbf{C} shows the operating points for each controller on top of the calibrated curve. Axis are as in B. Error bars on the marker show the standard error of mean across the 2000 testing samples.}
    \label{fig:control_calibration}
\end{figure}

Figure~\ref{fig:control_calibration} A is an example showing one episode of how aDBS control through a PPO controller is simulated forward in decision intervals of 250\,ms. At each decision point (red markers), the controller did regulate the action value, which is defining the mixture between the $\mathbf{\Theta}^{\text{OFF}}$ and $\mathbf{\Theta}^{\text{ON}}$ models that is used to simulate forward.
Using energy for DBS comes at a cost. In a best case scenario, this is just the energy consumed by the stimulator. More realistically, this cost would also be associated with side effects which are induced by DBS, which is difficult to quantify reliably in short time intervals. We calibrated cost and reward in the simulation environment, such that continuous DBS (cDBS), i.e., simulating with $\mathbf{\Theta}^{\text{ON}}$ only, theoretically nets a reward of zero. Various other cost values are shown in the calibration curves in Figure~\ref{fig:control_calibration} B. The panel shows what the average reward would be given a fixed action a. It is equivalent to cDBS with a fraction of the maximum DBS amplitude, where maximum means the value used during data recording. The calibration curves show that with small or zero cost, any amount of stimulation leads to a score improvement. While for high costs a mean reward optimization can be achieved by reducing the amplitude of cDBS (up to $c=1.6$), or no reward better than the all OFF case can be achieved. Both extremes are ill defined for the objective of training a controller for aDBS. The selected $c=1$ is the last of the tested values in which no optimization is achieved by simply choosing a cDBS with reduced amplitude. The maximum of the selected curve is not at zero as even an all-ON strategy is only creating 84.7\,\% episodes classified as DBS ON.
The operating points of the different control strategies including the static all-OFF or all-ON (cDBS with $a=1$) strategies are shown in Figure~\ref{fig:control_calibration} C. It shows that all aDBS controllers operate at a high average activation but surpass the all-ON strategy on average.

\begin{figure}
    \centering
    \includegraphics[width=\linewidth]{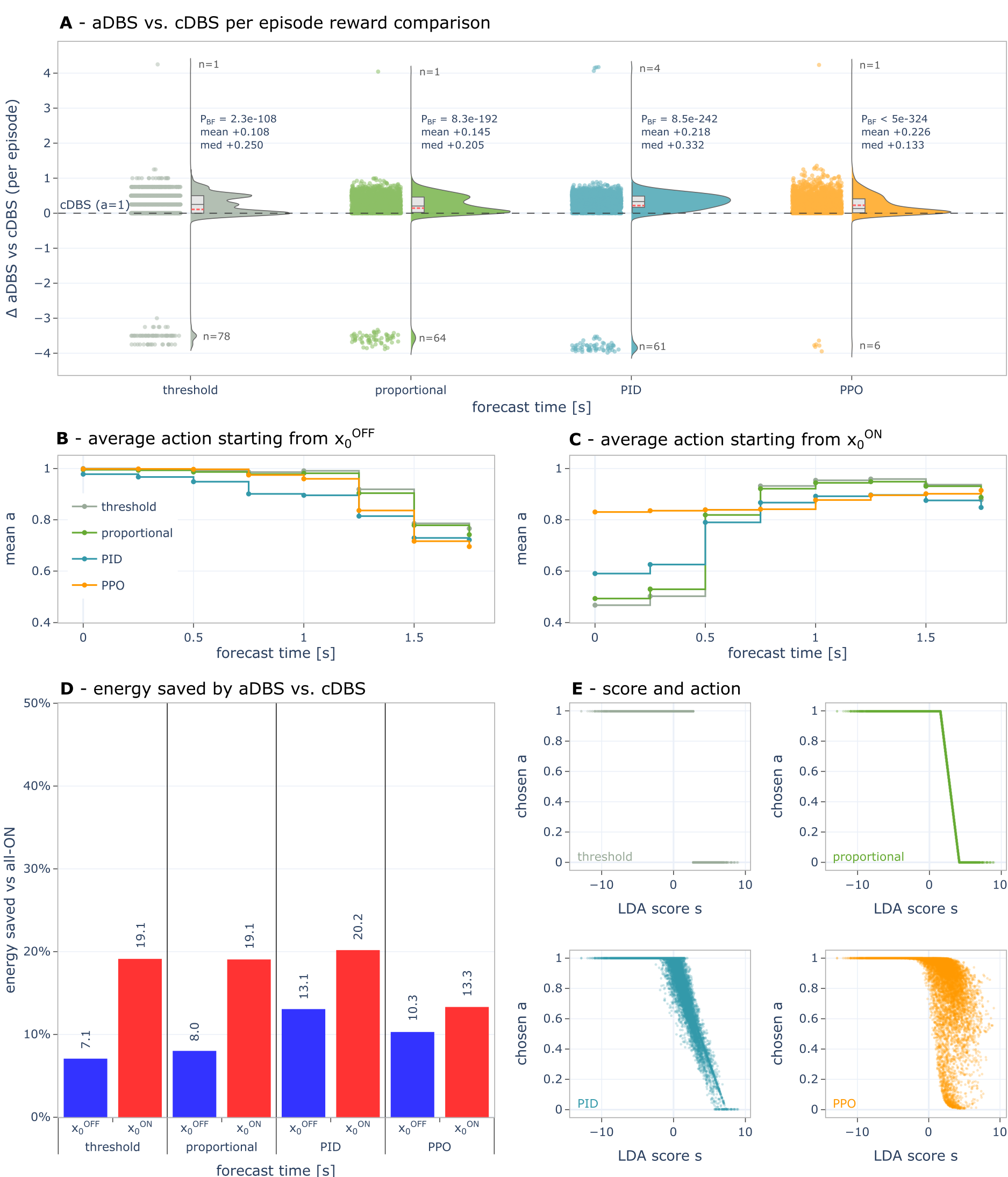}
    \caption{Control strategy comparison. Panel \textbf{A} shows paired comparisons between the aDBS controllers and the all-ON cDBS. The y-axis shows the reward $\Delta$. For each aDBS controller, individual markers correspond to the paired reward delta on the 2000 sample testing set. Half-violins show the kernel density estimation, with additional box plots for the inter quartile range, the mean (red dashed line) and the median (black line). Statistics show two-sided Wilcoxon signed-rank test of the trial-averaged paired differences against zero ($n=144$ trials), Bonferroni-corrected over the four controllers. Panel \textbf{B} and \textbf{C} show the average action value (y-axis) along the episode time separately for episodes starting with $\mathbf{x}_0^{\text{OFF}}$ (initial value drawn from DBS OFF data) and $\mathbf{x}_0^{\text{ON}}$ (initial value drawn from DBS ON). Panel \textbf{D} shows \textit{energy savings} compared to continuous all-ON (a=1) DBS, by fraction of the action of each aDBS controller. Savings are color coded according to the DBS condition of the initial value (blue for DBS OFF, red for DBS ON). Panel \textbf{E} shows a scatter plot with LDA decision score on the x-axis and the controller chosen action on the y-axis. One facet for each of the four aDBS controllers. An individual scatter point corresponds to a decision point ($\Delta t=250\,\text{ms}$, eight decision points per episode).}
    \label{fig:controller_performance}
\end{figure}

A paired comparison between the different aDBS controllers and the all-ON strategy reveals that all aDBS controllers have a significant advantage (Wilcoxon signed-rank test on trial-averaged paired reward differences, $n=144$ trials, $\text{P}_{BF}<10^{-8}$ for each controller; Figure~\ref{fig:controller_performance}~A). The lower tail of the distributions shows cases in which a cDBS achieved a simulation which was classified as DBS ON while the aDBS failed to do so. PPO has the least cases in which it falls short of cDBS (6 out of 2000). While having the largest mean effect size as well, the distributional shape shows, that the mass is mainly centered around zero (on par with cDBS, confirmed by the small median=0.133), with a positively skewed tail. Both the \textit{proportional} and PID control have very similar distributions, indicating that PID mainly learned a proportional control, which is confirmed by Figure~\ref{fig:controller_performance} E. \textit{Threshold} control has a binary action space which paired with the discrete and finite decision points lead to discrete reward levels. On the positive ends appear n=1 and n=4 (for PID) episodes for which the paired comparison is larger than the plain +2 reward. These are episodes for which the cDBS forecast was classified as DBS \OFF, while the aDBS controlled forecast was classified as DBS \ON. All episodes can be considered as outliers with decision score just marginally surpassing s>0.

Figure~\ref{fig:controller_performance} B and C show when the controllers decide to stimulate, depending on whether the initial value came from a DBS OFF ($\mathbf{x}_0^{\text{OFF}}$) or DBS ON ($\mathbf{x}_0^{\text{ON}}$) regime. Starting from DBS OFF data (Figure~\ref{fig:controller_performance} B), all controllers start at a high average action value (mainly DBS ON), while backing off and reducing DBS at about 1\,s into the simulation. PID and PPO reduced stimulation more strongly than \textit{threshold} and \textit{proportional} control (see also the energy savings for $\mathbf{x}_0^{\text{OFF}}$ in Figure~\ref{fig:controller_performance}~D).
When departing from DBS ON data (Figure~\ref{fig:controller_performance} C), a clear difference between PPO and the other control types is visible. While the other controllers start off by reducing stimulation to about 50\,\% of the action range, PPO never drops below 80\,\%. This more conservative start allows PPO maintaining a more stable low stimulation throughout, while the other controllers uniformly increase stimulation with a short rise at $500\,\text{ms}$, action values higher than PPO's between $750\,\text{ms}$--$1500\,\text{ms}$ on average. All controllers achieve higher energy savings (compared to all-ON cDBS) for episodes starting from $\mathbf{x}_0^{\text{ON}}$ than for episodes starting from $\mathbf{x}_0^{\text{OFF}}$ (Figure~\ref{fig:controller_performance} D). Overall, savings range from $7.1\,\%$ for \textit{threshold} on $\mathbf{x}_0^{\text{OFF}}$ to $20.2\,\%$ on $\mathbf{x}_0^{\text{ON}}$ for PID control. The difference between the DBS conditions of the initial values is lowest for PPO (most balanced saving, 10.3\,\% vs. 13.3\,\%), while the PID control surpasses each other controller in energy savings for initial values from both DBS conditions.
PID and PPO do not differ in reward (trial-level paired difference $-0.010$, Wilcoxon signed-rank test, $\text{P}_{BF}=1.0$), while PID uses significantly less energy (mean action $0.049$ lower, $\text{P}_{BF}<10^{-10}$). \textit{Proportional} control achieves a small but significant reward gain over \textit{threshold} control ($+0.037$, $\text{P}_{BF}=0.003$) at comparable energy use ($\text{P}_{BF}=0.052$).
During the simulation, the LDA decision score ($s$) in the 500\,ms lookback window was used as the physiomarker. Figure~\ref{fig:controller_performance} E relates this score at every decision point of each episode to the controllers action. It shows that the optimal threshold is at a positive bias with $s=2.8$ and not at $s=0$ (the LDA's decision boundary). Comparing the \textit{proportional} and PID control, it is visible that PID mostly controls in a proportional way, while differences are visible especially for smaller values of the score $s$. A more complex dependency is visible for PPO. A remarkable difference can be seen in that for many decision points with high LDA decision scores ($s>5$), the controller would maintain stimulation while others would reduce close to zero for $s>5$.

\section{Discussion}
In this work we demonstrate how system identification can be utilized to create a simulation environment for training aDBS controllers. This is a novel data-driven approach on tuning control parameters for aDBS using the system identification of nonlinear dynamics (SINDy) framework. Similar to previous data-driven simulation approaches~\cite{Wang2017,Yang2018} it bridges between existing model-based simulations using (single) neuron models and the actual human-in-the-loop system. A central advantage lies in its versatility. The approach can be used with any labeled biomarker recordings, regardless of the signal modality.
This work introduces the approach and demonstrates feasibility using robust ensemble SINDy (E-SINDy) on data from a single subject.
Based on this real patient EEG data, we applied methods from brain-computer interfacing, to extract a low dimensional signals with high contrast between the DBS conditions. These signals were then used with E-SINDy to fit two distinct models, one per DBS condition (\ON, \OFF). Numerical integration is used to forecast potential future data from these models. The forecasting capacity enables a \textit{what-if} analysis, estimating the brain's reaction to a change in DBS. Based on classification of the forecasts, we used Bayesian optimization and reinforcement learning to tune threshold, proportional, PID and PPO control strategies.

\subsection{System identification}
\subsubsection{Model selection}
Using E-SINDy we were able to predict narrowband signals over a future horizon ($\tau=2\,\text{s}$), while maintaining DBS specific dynamics.
The model fitting process reflects a general trade-off between model stability, accuracy, and expressiveness~\cite{Mangan2017, Fasel2022}---a version of the classical bias-variance trade-off~\cite {Belkin2019}. We selected model hyperparameters based on the accuracy of predicting the envelope of the future 2\,s of data, quantified by the eRMSE, and the ability to predict the training data domain of each model based on the forecasts they produce. The latter is quantified by the balanced recovery.
Interestingly, the models with the highest balanced recovery were models which only contained linear terms, but with otherwise the least regularization (Figure~\ref{fig:threshold_clip_grid} C). Linear models (of higher order) might hence be sufficient to capture the signal differences caused by DBS, which is a prerequisite for our approach. This links to linearized systems which have been proposed in the context of broader adaptive neurostimulation by means of model predictive control and linear quadratic regulation~\cite{Ahmadipour2021, Liang2023, Kia2025}.

\subsubsection{Characteristics of the selected models}
Our forecasts captured the dynamics of the true future data (Figure~\ref{fig:forecasting_results}) and produced a comparable power distribution over the 14--18\,Hz band through three distinct oscillatory modes (Figure~\ref{fig:threshold_clip_grid} G-L). Since the spacing of the modes is not perfectly symmetric (see eigenvalues in Figure~\ref{fig:threshold_clip_grid}), the resulting amplitude modulation is quasi-periodic only---a property which distinguishes a burst-like envelope from a periodic (stationary) beat. However, the eigenvalues (Figure~\ref{fig:threshold_clip_grid}, H and K) and the frequencies of the according eigenmodes (appendix Table~\ref{xtab:eigenmodes} create multiple beat periods in the range of 331--769\,ms, see appendix Table~\ref{xtab:beatperiods}. These durations are well within the range of reported physiological beta bursts, which have been proposed as biomarkers for aDBS~\cite{Tinkhauser2017, Tinkhauser2017b, he:2023}. Additionally, our forecast examples (Figure~\ref{fig:forecasting_results}) suggest that it might be possible to predict a beta burst onset much further in advance than previously reported 80--100\,ms ~\cite{Abdi-Sargezeh2025}, if bursts are defined as power beyond a given quantile of past data~\cite{Tinkhauser2017, Tinkhauser2017b, he:2023}.
Our models are able to create a distribution rather than a fixed set of beats, because the beat times of the mode differences (appendix Table~\ref{xtab:eigenmodes}) do not synchronize (no common factors). Analyzing this burst distribution is outside the scope of this work, but we note that for $\mathbf{\Theta}^{\text{ON}}$ model, the maximal time constant ($T_{beat}=715.8\,\text{ms}$) is faster than for the $\mathbf{\Theta}^{\text{OFF}}$ ($T_{beat}=768.9\,\text{ms}$)---a potential indication for shorter beta bursts under DBS ON, which agrees with existing literature ~\cite{Tinkhauser2017, Tinkhauser2017b}.

\subsubsection{DBS differences and switch dynamics}
Both models ($\mathbf{\Theta}^{\text{OFF}}$, $\mathbf{\Theta}^{\text{ON}}$) capture dynamics inherent to their training data regime (DBS \ON, DBS \OFF). This becomes apparent in the significantly larger eRMSE for the counterfactual forecasts (Figure~\ref{fig:forecasting_results}, D and J). The analysis of the classification results and the bandpower of the two channels' forecasts (Figure~\ref{fig:forecast_classification}) show that the models are able to drive dynamics to a regime resembling their training data distribution within approximately 1\,s and 1.5\,s for $\mathbf{\Theta}^{\text{OFF}}$ and $\mathbf{\Theta}^{\text{ON}}$ respectively.
While there is no precise timing correspondence with existing literature, the range of this dynamic change is plausible but falls within a time window which is not closely studied for wash-in/wash-out effects. On the shorter end, DBS evoked potentials are known to rapidly change in the first 100\,ms after DBS onset and keep changing over a 60\,s window with strongest amplitudes at around 3\,s~\cite{Schmidt2020}. Beta bandpower suppression in local field potentials starts at 500\,ms after DBS onset~\cite{Wiest2020}, while cortical adjustments are observed in magnetoencephalography in the first ~250\,ms after DBS onset~\cite{Spooner2024}. For longer lasting effects, time resolution is often in the 10\,s~\cite{Bronte-Stewart2009, Wiest2021}, with behavioral wash-out effects being in the order of 10s of minutes~\cite{Cooper2013}.
The relevance of this time frame, and thus the plausibility of our simulation timing, is additionally supported by the finding that DBS delivered in bursts as short as 300\,ms improves performance in a hand-motor task~\cite{Cavallo2025}, the same motor domain as our data.

\subsection{Control simulation environment}
Using both models, we created a simulation environment which incorporates a control action $\text{a}\in \left[0, 1\right]$ via a linear mixture of the two models.
The controller defines the action based on past data (LDA decision score on bandpower of the last 500\,ms). The reward is calculated based on the classification of a full 2\,s simulation window, and is reduced by the energy cost of the actions taken.

For this proof of concept, we chose a neutral cost calibration (cDBS nets no reward). A personalized calibration could account for person-specific side effect levels, such that more severe side effects would incur a higher cost. Our calibration curves show that with a higher cost value, a reward optimization can already be achieved with a modest reduction in cDBS amplitude (Figure~\ref{fig:control_calibration} B). Such a \textit{cost free optimization} can be plausible as the therapeutic window is not a single stimulation amplitude~\cite{Smeets2024}.  A small amplitude reduction (e.g., ~10\,\%) could maintain similar effectiveness as the original amplitude while strictly consuming less energy. Our calibration was chosen to rule out such a cost-free optimization for our simulation environment.

With the given cost calibration and episodes of 2\,s, all controllers operate close to the all-ON cDBS in average action, i.e. stimulation intensity or current (Figure~\ref{fig:control_calibration} C). However, all controllers were able to exploit state-dependent stimulation, with improvements beyond those obtainable by a mere general reduction of the stimulation intensity. The \textit{when} to stimulate matters.

Another advantage of the simulation environment is that it allows us to create paired comparisons using identical initial conditions. Figure~\ref{fig:controller_performance} A shows that all controllers (aDBS) are able to achieve significant reward improvements compared to the all-ON cDBS.

Overall, energy savings are modest with 7.1\,\% to 13.1\,\% for initial values from the DBS \OFF condition ($\mathbf{x}_0^{\text{OFF}}$), and 13.3\,\% to 20.2\,\% for initial values from the DBS \ON condition ($\mathbf{x}_0^{\text{ON}}$)---see Figure~\ref{fig:controller_performance} D. A \textit{threshold} control already realized most of the energy saving potential. The most efficient control is delivered by the PID controller with energy savings of 13.1\,\% for $\mathbf{x}_0^{\text{OFF}}$ and 20.2\,\% for $\mathbf{x}_0^{\text{ON}}$, and a mean paired effect over cDBS of +0.218. The energy savings of our controllers are smaller than previously reported values of~$50\,\%$ reductions in experimental settings~\cite{Little2013, Velisar2019, Bocci2021}. However, our savings align well with results from a recent large cohort with long-term personalized aDBS, reporting 15--19\,\% reduction of total electrical energy delivered~\cite{brontestewart2025}. %

While the energy or action efficiency is a read-out for our simulation approach, it should not be considered in the light of energy savings alone. As mentioned above, cDBS can lead to side effects, while aDBS is reported in some cases to improve symptoms beyond cDBS~\cite{Little2013, Rosa2015, Bocci2021}. The question of \textit{when to stimulate} therefore extends beyond pure energy concerns. Our approach would need to be extended to include such symptomatic information to reflect additional cost or benefits and optimize for the trade-off.

The current approach already enables the comparison of DBS timing dependency and the overall (re)action profile of controllers to the physiomarker (Figure~\ref{fig:controller_performance} B, C, E).
We note that the optimized \textit{threshold} and \textit{proportional} control are very similar as a result of the dynamic range of the \textit{proportional} controller being small. This can be attributed to the Bayesian optimization, and is not a free choice. Given our approach, we can conclude that both controllers have a similar performance, in line with similar clinical effects being reported for implanted systems~\cite{Arlotti2021, Bocci2021}. The control strategies PID and PPO have a richer set of parameters to optimize. Similar to a previous comparison of threshold and PI(D) control in simulated data~\cite{Fleming2023}, PID is the most efficient controller in our simulation setting. It achieves this efficiency especially by reducing stimulation in $\mathbf{x}_0^{\text{OFF}}$ episodes earlier than other controllers (Figure~\ref{fig:controller_performance} B).

PPO did not surpass PID in reward (no significant difference) and used more energy, although it has the capacity to learn more complex control patterns due to its more complex parameters. For a fair comparison, we only provided the LDA decision score and CSP log-bandpower as state information for the RL environment. It is possible that PPO could optimize the control further, if richer state information was provided. PPO followed a more conservative strategy by keeping the stimulation high when starting from $\mathbf{x}_0^{\text{ON}}$ data (Figure~\ref{fig:controller_performance} C), and keeping stimulation high, when the physiomarker was already indicating an \textsc{ON}-like state (Figure~\ref{fig:controller_performance} E). This resulted in a more stable stimulation, and as a consequence, PPO did save the least energy on average (mean of DBS ON and OFF savings, Figure~\ref{fig:controller_performance} D). However, it fell short of cDBS in the least number of episodes (n=6 (0.3\,\%), Figure~\ref{fig:controller_performance} A) compared with all other controllers (n>60). This highlights another potential asymmetry which could be included in future iterations of the simulation environment, e.g., using an additional cost to model the requirement to perform at least on the level of cDBS. Such a requirement would reflect the clinical importance of the DBS effect.

From the controller comparison, we can conclude that within our simulation environment, PID is the preferable control strategy:  its reward advantage over cDBS is statistically indistinguishable from that of PPO, while it uses the least energy of all controllers (significantly less than PPO). Our results suggest that at least until more complex physiomarker signals become available, the investment in deep RL control might not pay off in efficiency gains.

\subsection{Limitations and future work}
Our proposed approach and the proof-of-concept validation span a wide range of theoretical concepts, including signal processing, system identification and control theory. The purpose of this manuscript is to introduce the approach with the objective of creating a data-driven, personalized simulation environment. Additionally, we chose to provide a rich analysis of the fitted models' suitability, rather than benchmark the whole approach. Each stage of the pipeline---physiomarker extraction, system identification, and controller optimization---offers its own optimization potential and could warrant dedicated studies. We therefore deliberately limited this work to demonstrating the feasibility of the complete pipeline and deferred a systematic benchmark across subjects and model variants to future work.

Regardless, we acknowledge the limitation of the proof-of-concept being evaluated with a single hand-picked subject. The selection was based on previous results demonstrating a strong behavioral and electrophysiological effect of DBS for this subject (S2, session 2)~\cite{Dold2026}. It remains to be explored how the approach performs on less favorable data.

With less favorable data, the question arises of how model fitting could be further optimized. We chose a robust but otherwise plain (STLSQ optimization) version of the E-SINDy approach. The idea of using system identification for creating a control simulation environment is general and does not depend on a SINDy-type model~\cite{Yang2018}. Future research might explore different system identification methods~\cite{Ahmadipour2021, Liang2023, Kia2025}, or improved versions of SINDy~\cite{Zheng2019}. Our proof-of-concept results suggest that linear models might be sufficient in the trade-off between eRMSE and balanced recovery.

In terms of signals, we limited our data to two CSP channels from a single frequency band. Future work should explore multi-channel and multi-band data, as system identification would lend itself naturally to incorporating cross-frequency coupling, which was shown to be modulated under DBS~\cite{Muthuraman2020}.
Additionally, different preprocessing pipelines might improve the signal further, e.g., using dynamic mode decomposition~\cite{Fukuma2024}.

\begin{figure}
    \centering
    \includegraphics[width=\linewidth]{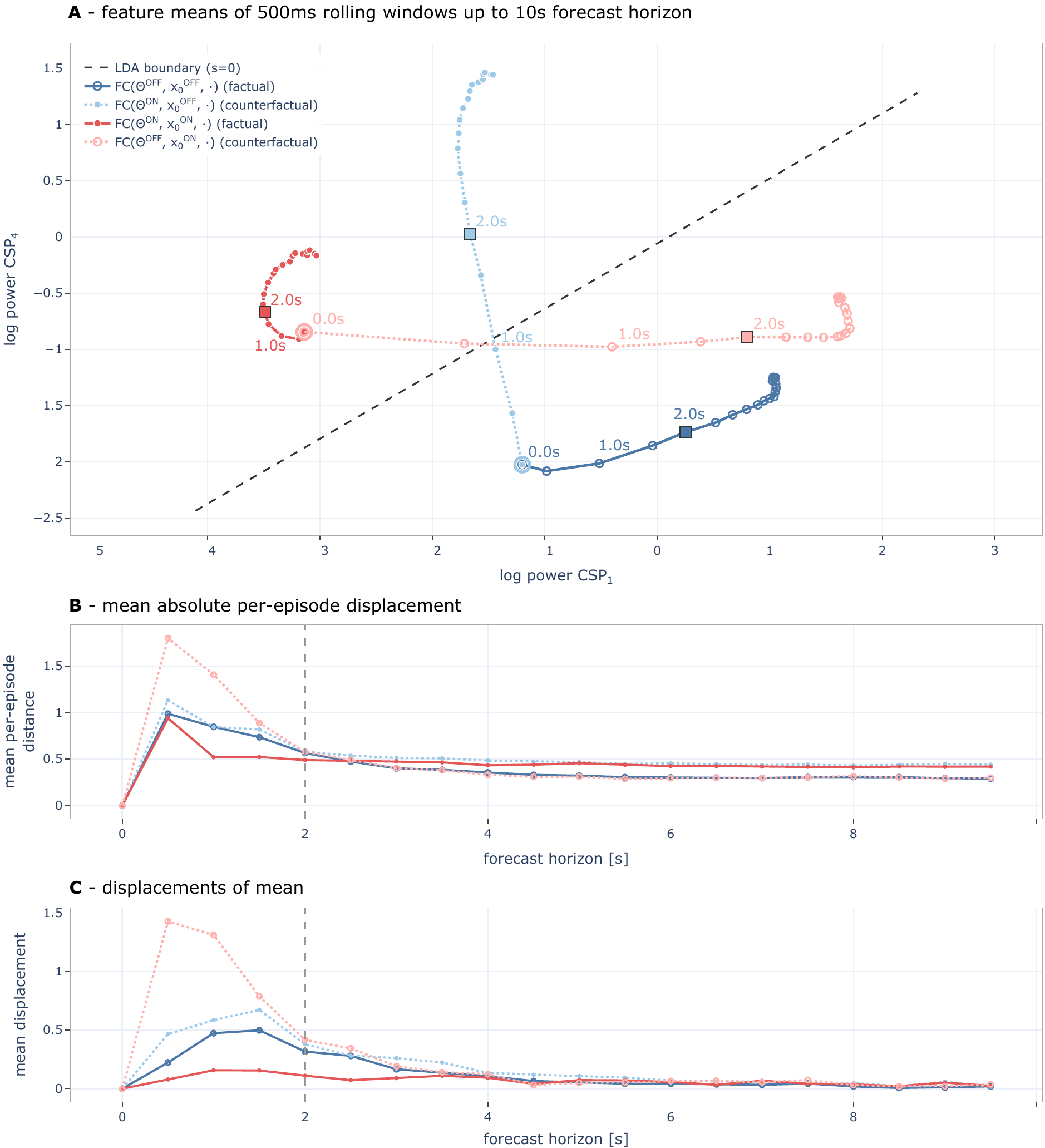}
    \caption{Bandpower estimation in 500\,ms rolling windows for forecasts up to 10\,s into the future. 600 DBS ON and 600 DBS OFF starting positions ($\mathbf{x}_0^{\text{ON}}$,$\mathbf{x}_0^{\text{OFF}}$ ) were drawn randomly from the 4--6\,s interval of each of the 144 trials. Forecasts were created for both models ($\mathbf{\Theta}^{\text{ON}}$, $\mathbf{\Theta}^{\text{OFF}}$). For each forecasted series, the log power was estimated every 250\,ms over a sliding window of 500\,ms in size (50\,\% overlap).}
    \label{xfig:500ms_rolling_window_estimation}
\end{figure}

Another limitation is the short forecasting horizon (2\,s), which we chose as a trade-off between maintaining realistic characteristics due to starting from real data initial values, while allowing enough time for the DBS regime-specific dynamics to manifest. While we analyzed the bandpower features on forecasts up to 10\,s into the future (appendix Figure~\ref{xfig:500ms_rolling_window_estimation}), additional research is required to understand longer control horizons. This includes the question if longer horizons add valuable information at all.

Similar to the related data-driven approaches~\cite{Yang2018, Yang2021, Haddock2017, Wang2017}, the evaluation of the control strategies happened in simulation only. Whether the performance differences are therapeutically meaningful can only be established with the human-in-the-loop, which is the ultimate validating any simulation environment would need to be measured against.

Finally, regarding the control simulation environment and control algorithms, there are a couple of choices which---while plausible first steps in our opinion---should be investigated further. For the simulation environment, this specifically concerns the linear mixture of the $\mathbf{\Theta}^{\text{OFF}}$ and $\mathbf{\Theta}^{\text{ON}}$ dynamics. While a monotonic effect of increased stimulation up to the therapeutic window is observable with common beta power physiomarkers~\cite{Feldmann2022}, it is unclear whether the transition between DBS regimes would follow the simplified linear mixture we proposed. Using, e.g., a sigmoidal mixture could be another plausible choice, which, however could introduce a potential effect over cDBS purely from the mixture parameters themselves. We still consider a linear mixture a plausible first candidate, as it directly links to a pure proportional control that is usable in certified devices~\cite{Artoni2018}, but other mixtures, potentially time or stimulation dependent themselves could be calibrated with more data.
With respect to the RL controllers, we note that PPO is just one particular representative which we used with mainly default parameters (StableBaseline 3). Future work might investigate other RL-type controllers and PPO configurations tuned for the aDBS control task, in addition to considering a richer signal (more channels, additional modalities, or preprocessed features) as we mentioned above.

\section{Acknowledgments}

MD receives funding from the Dareplane collaboration project, which is co-funded by PPP Allowance awarded by Health~Holland, Top Sector Life Sciences \& Health, to stimulate public-private partnerships as well as by a contribution from the Dutch Brain Foundation. The PD-Interaktiv I study is funded by the Bundesministerium für Bildung und Forschung (Grant 16SV8011).
JP is funded by the Deutsche Forschungsgemeinschaft (DFG) under the Walter Benjamin Program (Project number 510112977). The other authors declare not to have conflicting interests.

\section{Author Contributions}
All authors: Writing - Review and Editing. MD: Conceptualization, Methodology, Software, Validation, Formal Analysis, Investigation, Data Curation, Writing - Original Draft, Visualization, Project Administration.  JP: Conceptualization, Methodology, Investigation, and Project Administration, Writing - Original Draft. MLFJ: Conceptualization, Investigation, Resources, Writing - Original Draft, Supervision, Project Administration, Funding Acquisition. MT: Conceptualization, Methodology, Investigation, Resources, Writing - Original Draft, Supervision, Project Administration, Funding Acquisition.

\bibliographystyle{vancouver}   %
\bibliography{bibliography} %

\section{Appendix}

\begin{figure}
    \centering
    \includegraphics[width=\linewidth]{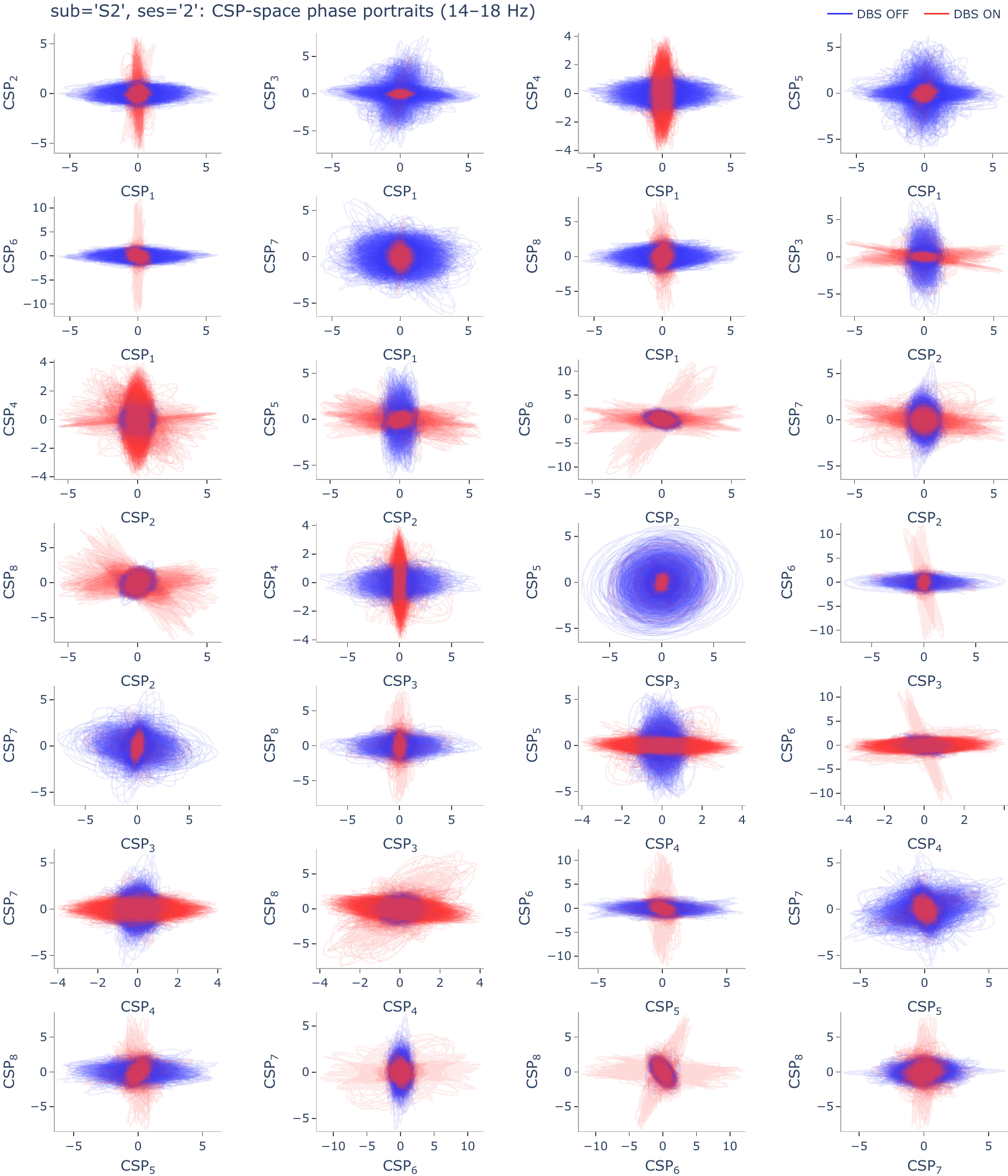}
    \caption{Common spatial patterns (CSP) bivariate trajectory phase spaces for each pair of CSP components in the 14--18\,Hz band, one trajectory per trial, colored by DBS condition. The pair ($\text{CSP}_1$, $\text{CSP}_4$) showing the clearest and most stable separation was selected for modeling.}
    \label{xfig:csp_phase_space}
\end{figure}

\begin{figure}
    \centering
    \includegraphics[width=\linewidth]{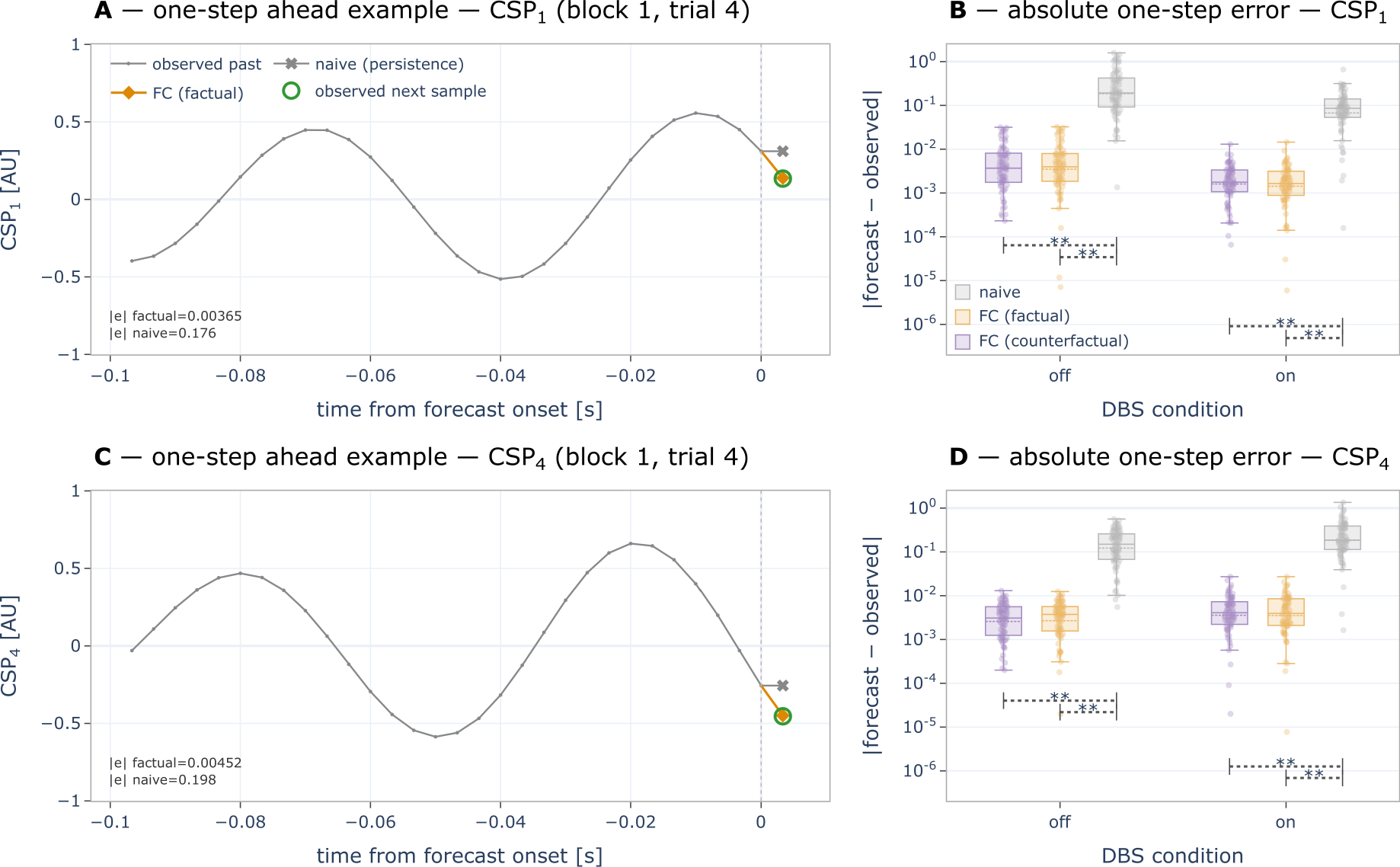}
    \caption{One-step ahead forecasts. Absolute error of the factual model's one-step prediction compared with a naive steady-state (persistence) prediction on the same trials. Model predictions are significantly more accurate for both channels and both DBS conditions (two-sided Wilcoxon signed-rank test, $n=72$ trials each, $\text{P}_{BF}<10^{-12}$, Bonferroni correction over the four channel-condition combinations).}
    \label{xfig:one_step_forecasts}
\end{figure}

\begin{table}[htbp]
  \centering
  \caption{Eigenmodes of the identified models $\mathbf{\Theta}^{\text{OFF}}$ and $\mathbf{\Theta}^{\text{ON}}$. Each row is one complex-conjugate eigenvalue pair $\lambda = \sigma \pm 2\pi i f$; the $12$-dimensional state gives six pairs per model. The pairs fall into three spectral clusters of two, each cluster holding one damped pair and one pair resting on the stabilizer floor. $\tau = -1/\sigma$ is the amplitude time constant: the mode's envelope $e^{\sigma t}$ decays to $1/e \approx 37\%$ of its initial value after $\tau$.}
  \label{xtab:eigenmodes}
  \begin{tabular}{clrrrr}
    \toprule
    \# & cluster & $f$ [Hz] & $T_{\mathrm{cyc}}$ [ms] & $\sigma$ [s$^{-1}$] & $\tau$ [s] \\
    \midrule
    \multicolumn{6}{l}{\emph{$\mathbf{\Theta}^{\mathrm{OFF}}$}} \\
    1 & lower & 14.2493 & 70.18 & -0.00001 & $\ge 10^{5}$\rlap{$^{\dagger}$} \\
    2 & lower & 14.3658 & 69.61 & -0.06397 & 15.63 \\
    3 & central & 15.9228 & 62.80 & -0.45257 & 2.21 \\
    4 & central & 15.9390 & 62.74 & -0.00001 & $\ge 10^{5}$\rlap{$^{\dagger}$} \\
    5 & upper & 17.2395 & 58.01 & -0.00001 & $\ge 10^{5}$\rlap{$^{\dagger}$} \\
    6 & upper & 17.2665 & 57.92 & -0.60878 & 1.64 \\
    \addlinespace
    \multicolumn{6}{l}{\emph{$\mathbf{\Theta}^{\mathrm{ON}}$}} \\
    1 & lower & 14.2530 & 70.16 & -0.15715 & 6.36 \\
    2 & lower & 14.3108 & 69.88 & -0.00001 & $\ge 10^{5}$\rlap{$^{\dagger}$} \\
    3 & central & 15.8209 & 63.21 & -0.29946 & 3.34 \\
    4 & central & 15.8451 & 63.11 & -0.00001 & $\ge 10^{5}$\rlap{$^{\dagger}$} \\
    5 & upper & 17.2421 & 58.00 & -0.15711 & 6.37 \\
    6 & upper & 17.2527 & 57.96 & -0.00001 & $\ge 10^{5}$\rlap{$^{\dagger}$} \\
    \bottomrule
  \end{tabular}
  \par\vspace{2pt}
  \footnotesize $^{\dagger}$Pair sits on the stabilizer floor ($\sigma = -\epsilon_{\mathrm{clip}} = -10^{-5}$). The implied $\tau$ is not identified persistence, only the absence of further clipping.
\end{table}

\begin{table}[htbp]
  \centering
  \caption{Envelope modulation periods $T = 1/\Delta f$ implied by every pairwise frequency difference among the six eigenmodes of Table~\ref{xtab:eigenmodes}. Superposing two modes $\Delta f$ apart beats at this period, independent of whether the modes are physically coupled.}
  \label{xtab:beatperiods}
  \begin{tabular}{llrr}
    \toprule
    pair & & $\Delta f$ [Hz] & $T_{\mathrm{beat}}$ [ms] \\
    \midrule
    \multicolumn{4}{l}{\emph{$\mathbf{\Theta}^{\mathrm{OFF}}$}} \\
    1--6 & & 3.0172 & 331.4 \\
    1--5 & & 2.9902 & 334.4 \\
    2--6 & & 2.9007 & 344.7 \\
    2--5 & & 2.8737 & 348.0 \\
    1--4 & & 1.6897 & 591.8 \\
    1--3 & & 1.6735 & 597.5 \\
    2--4 & & 1.5732 & 635.6 \\
    2--3 & & 1.5570 & 642.2 \\
    3--6 & & 1.3437 & 744.2 \\
    4--6 & & 1.3275 & 753.3 \\
    3--5 & & 1.3167 & 759.5 \\
    4--5 & & 1.3005 & 768.9 \\
    1--2 & & 0.1165 & 8.6\,s\rlap{$^{\ddagger}$} \\
    5--6 & & 0.0270 & 37.1\,s\rlap{$^{\ddagger}$} \\
    3--4 & & 0.0162 & 61.7\,s\rlap{$^{\ddagger}$} \\
    \addlinespace
    \multicolumn{4}{l}{\emph{$\mathbf{\Theta}^{\mathrm{ON}}$}} \\
    1--6 & & 2.9997 & 333.4 \\
    1--5 & & 2.9891 & 334.6 \\
    2--6 & & 2.9420 & 339.9 \\
    2--5 & & 2.9313 & 341.1 \\
    1--4 & & 1.5921 & 628.1 \\
    1--3 & & 1.5678 & 637.8 \\
    2--4 & & 1.5343 & 651.7 \\
    2--3 & & 1.5101 & 662.2 \\
    3--6 & & 1.4319 & 698.4 \\
    3--5 & & 1.4212 & 703.6 \\
    4--6 & & 1.4076 & 710.4 \\
    4--5 & & 1.3970 & 715.8 \\
    1--2 & & 0.0577 & 17.3\,s\rlap{$^{\ddagger}$} \\
    3--4 & & 0.0242 & 41.2\,s\rlap{$^{\ddagger}$} \\
    5--6 & & 0.0107 & 93.9\,s\rlap{$^{\ddagger}$} \\
    \bottomrule
  \end{tabular}
  \par\vspace{2pt}
  \footnotesize $^{\ddagger}$Difference between two pairs of the same spectral cluster; listed for completeness.
\end{table}

\begin{figure}
    \centering
    \includegraphics[width=\linewidth]{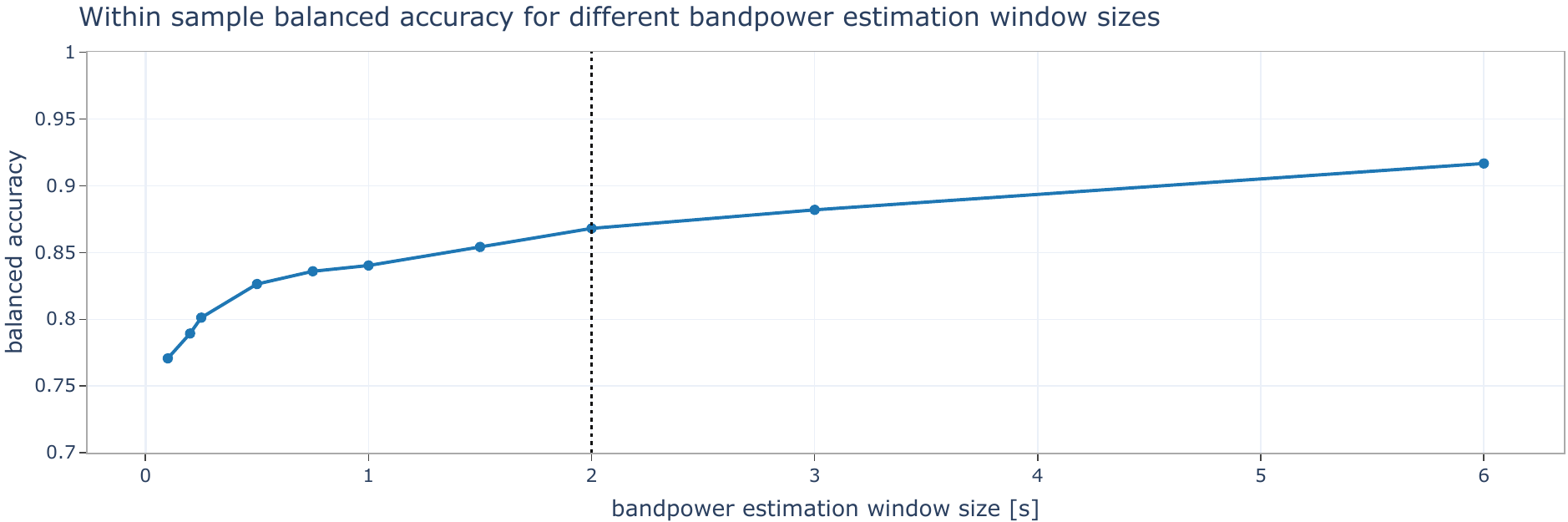}
    \caption{Bandpower estimation in various window sizes and according balanced accuracy for an LDA classifier trained on full 6\,s of all 144 available trials. For window sizes smaller than 6\,s, each trial was separated in non-overlapping windows of that size covering the full trial (each of the tested windows dividing 6 without remainder). Then balanced accuracy is calculated based on the classification of each of the smaller windows. The evaluation is fully in-training-sample, as the objective is to understand when the bandpower estimation with respect to the final LDA stabilizes.}
    \label{xfig:bandpower_window_sizes_vs_balanced_acc}
\end{figure}

\end{document}